\documentclass[10pt,reqno,a4paper]{amsart}
\usepackage[british]{babel}
\usepackage{amssymb,amstext,amsthm,eucal,bbm,mathrsfs,amscd,bbold}
\usepackage[margin=2cm]{geometry}
\usepackage{array}
\usepackage[svgnames]{xcolor}
\usepackage[utf8]{inputenc}
\usepackage{enumitem}
\usepackage[small]{eulervm}
\usepackage{tgpagella}
\usepackage{tikz,pgfplots}
\usetikzlibrary{calc}
\pgfplotsset{compat=1.14}
\usepackage[unicode]{hyperref}
\hypersetup{%
  pdftitle   = {Higher-dimensional kinematical Lie algebras via deformation theory},
  pdfkeywords = {Lie algebra, deformations, kinematical, central
    extension, higher dimensions},
  pdfauthor  = {José Figueroa-O'Farrill},
  pdfcreator = {\LaTeX\ with package \flqq hyperref\frqq},
  linkcolor=NavyBlue,
  citecolor=ForestGreen,
  urlcolor=OrangeRed,
  anchorcolor=OrangeRed,
  colorlinks=true
}
\theoremstyle{plain}

\theoremstyle{definition}

\renewcommand{\d}{\partial}
\newcommand{\g}{\mathfrak{g}}
\newcommand{\tg}{\tilde{\mathfrak{g}}}

\newcommand{\p}{\mathfrak{p}}
\newcommand{\co}{\mathfrak{co}}
\newcommand{\e}{\mathfrak{e}}

\newcommand{\so}{\mathfrak{so}}

\newcommand{\h}{\mathfrak{h}}
\renewcommand{\th}{\tilde{\mathfrak{h}}}
\newcommand{\s}{\mathfrak{r}}

\newcommand{\be}{\boldsymbol{e}}
\renewcommand{\t}{\boldsymbol{t}}

\newcommand{\R}{\boldsymbol{R}}
\newcommand{\B}{\boldsymbol{B}}

\newcommand{\bbeta}{\boldsymbol{\beta}}
\newcommand{\bpi}{\boldsymbol{\pi}}
\renewcommand{\P}{\boldsymbol{P}}

\renewcommand{\H}{\mathcal{H}}

\newcommand{\ad}{\operatorname{ad}}
\newcommand{\id}{\mathbb{1}}

\newcommand{\RR}{\mathbb{R}}

\newcommand{\Aff}{\operatorname{Aff}}
\newcommand{\GL}{\operatorname{GL}}
\newcommand{\SO}{\operatorname{SO}}

\definecolor{gris}{rgb}{0.8,0.8,0.8}
\newcommand{\zero}{{\color{gris}0}}
\allowdisplaybreaks[0]
\begin{document}

\title{Higher-dimensional kinematical Lie algebras via deformation theory}
\author{José M. Figueroa-O'Farrill}
\address{Maxwell Institute and School of Mathematics, The University
  of Edinburgh, James Clerk Maxwell Building, Peter Guthrie Tait Road,
  Edinburgh EH9 3FD, United Kingdom}
\email{j.m.figueroa@ed.ac.uk}
\begin{abstract}
  We classify kinematical Lie algebras in dimension $D+1$ for
  $D > 3$ up to isomorphism.  This is part of a series of papers
  applying deformation theory to the classification of kinematical Lie
  algebras in arbitrary dimension.  This is approached via the
  classification of deformations of the relevant static kinematical
  Lie algebra.  We also classify the deformations of the universal
  central extension of the static kinematical Lie algebra in dimension
  $D+1$ for $D>3$.  In addition we determine which of these Lie
  algebras admit an invariant inner product.
\end{abstract}
\thanks{EMPG-17-12}
\maketitle
\tableofcontents

\section{Introduction}
\label{sec:introduction}

In a previous paper \cite{JMFKinematical3D} we have presented an
approach to the classification of kinematical Lie algebras based on
deformation theory, extending earlier work \cite{JMFGalilean} for the
galilean and Bargmann algebras.  In \cite{JMFKinematical3D} we recovered
the classification of Bacry and Nuyts \cite{MR857383} of kinematical
Lie algebras  in dimension $3+1$, and extended it to classify also the
deformations of the universal central extension of the static
kinematical Lie algebra in that dimension.  The purpose of this paper
is to extend these classifications to dimension $D+1$ for all $D >
3$.  A separate paper \cite{TAJMFKinematical2D} will present the
classification of kinematical Lie algebras for $D=2$, which is
technically quite different than $D=3$ and $D>3$.  The results of this
series of papers is summarised in \cite{JMFKinematicalSummary}.

By a \textbf{kinematical Lie algebra} in dimension $D+1$, we mean a
real $\tfrac12 (D+1)(D+2)$-dimensional Lie algebra with generators
$R_{ab} = - R_{ba}$, with $1\leq a,b \leq D$, spanning a Lie
subalgebra $\s$ isomorphic to the Lie algebra $\so(D)$ of rotations in
$D$ dimensions; that is,
\begin{equation}
  [R_{ab}, R_{cd}] = \delta_{bc} R_{ad} -  \delta_{ac} R_{bd} -
  \delta_{bd} R_{ac} +  \delta_{ad} R_{bc},
\end{equation}
and $2D+1$ generators $B_a$, $P_a$ and $H$ which transform according
to the vector, vector and scalar representations of $\so(D)$,
respectively -- namely,
\begin{equation}
  \begin{split}
    [R_{ab}, B_c] &= \delta_{bc} B_a - \delta_{ac} B_b\\
    [R_{ab}, P_c] &= \delta_{bc} P_a - \delta_{ac} P_b\\
    [R_{ab}, H] &= 0.
  \end{split}
\end{equation}
The rest of the brackets between $B_a$, $P_a$ and $H$ are only subject
to the Jacobi identity: in particular, they must be $\s$-equivariant.
The kinematical Lie algebra where those additional Lie brackets vanish
is called the \textbf{static} kinematical Lie algebra.  Every other
kinematical Lie algebra is, by definition, a deformation of the
static one.

Up to isomorphism, there is only one kinematical Lie algebra with
$D=0$: it is one-dimensional and hence abelian.  For $D=1$, there are
no rotations and hence any three-dimensional Lie algebra is
kinematical.  The classification is therefore the same as the
celebrated Bianchi classification of three-dimensional real Lie
algebras \cite{Bianchi}.  As far as I know the only other
classification of kinematical Lie algebras is that in $3+1$
dimensions, by Bacry and Nuyts \cite{MR857383}; although there is also
a classification of kinematical superalgebras in $3+1$ dimensions
\cite{CampoamorStursberg:2008hm}.  The purpose of this paper is to
solve the classification problem for $D>3$ using deformation theory
along the lines of \cite{JMFGalilean,JMFKinematical3D}.  As we will
see, we can treat all $D>4$ in a uniform way (with the exception for
$D=5$ which is somewhat special but not in an essential way).  The
case $D=4$ is slightly more complicated due to $\so(4)$ being
semisimple but not simple.  Nevertheless as we will see, this case
reduces to the generic case ($D\geq 5$); although this requires a
calculation.  The case $D=2$ is computationally more involved because
$\so(2)$ is abelian and its vector representation (on $\RR^2$) has a
larger than normal endomorphism ring.  That case is the subject of a
separate paper \cite{TAJMFKinematical2D}.

The static kinematical Lie algebra in $D\geq 3$ admits a universal
central extension with central generator $Z$ and additional Lie
bracket
\begin{equation}
  [B_a, P_b] = \delta_{ab} Z,
\end{equation}
and we will also consider the problem of classifying the deformations
of the centrally extended static kinematical Lie algebra for $D>3$,
the case $D=3$ having been done in \cite{JMFKinematical3D}.

We refer to \cite{JMFKinematical3D} for the methodology and the basic
notions of deformation theory and Lie algebra cohomology as in
\cite{MR0214636,ChevalleyEilenberg,MR0054581}.

For applications to conformal field theory, gauge theory and even knot
theory, it is important for a Lie algebra to admit an invariant
symmetric inner product.  In this paper we also determine those
kinematical Lie algebras which admit such an inner product.
Disappointingly, perhaps, we find that only the simple kinematical Lie
algebras admit an invariant symmetric inner product.  Similarly, for
the extended kinematical Lie algebras, we will find that only the
trivial central extensions of the simple kinematical Lie algebras
admit such an inner product.

This paper is organised as follows.  In
Section~\ref{sec:deform-stat-kinem-1} we classify kinematical Lie
algebras in dimension $D+1$ for all $D\geq 5$, arriving at
Table~\ref{tab:summary}, while in Section~\ref{sec:kinem-lie-algebr}
we treat the case of $D=4$, but show, after some calculations, that we
obtain the same results as for $D\geq 5$.  Therefore
Table~\ref{tab:summary} is valid for $D\geq 4$.  Next, in
Section~\ref{sec:deform-centr-extend}, we determine the universal
central extension of the static kinematical Lie algebra and proceed to
classify its deformations for $D \geq 5$, arriving at
Table~\ref{tab:ce-summary}, while in
Section~\ref{sec:deform-centr-extend-4} we repeat the calculation for
$D=4$ arriving at the conclusion that Table~\ref{tab:ce-summary} also
holds for $D\geq 4$.  Comparison with the case of $D=3$ shows that
whereas in $D=3$ there are more kinematical Lie algebras which have no
analogue in $D>3$, the same is not true for the deformations of the
centrally extended algebra.  For those deformations, the results for
$D\geq 3$ are uniform.  Finally, in Section~\ref{sec:conclusions} we
offer some conclusions.

\section{Deformations of the static kinematical Lie algebra with $D\geq5$}
\label{sec:deform-stat-kinem-1}

Let $\g$ denote the static kinematical Lie algebra for $D \geq 5$ and
generators $R_{ab}$, $B_a$, $P_a$ and $H$, subject to the non-zero
brackets
\begin{equation}
  \begin{split}
  [R_{ab}, R_{cd}] & = \delta_{bc} R_{ad} -  \delta_{ac} R_{bd} - \delta_{bd} R_{ac} +  \delta_{ad} R_{bc}\\
  [R_{ab}, B_c] &= \delta_{bc} B_a - \delta_{ac} B_b\\
  [R_{ab}, P_c] &= \delta_{bc} P_a - \delta_{ac} P_b.
  \end{split}
\end{equation}
We often find it convenient to employ an abbreviated notation where
the indices are suppressed.  In that notation, we would write the
above brackets as
\begin{equation}
  \label{eq:static}
  [\R,\R] = \R  \qquad [\R, \B] = \B \qquad\text{and}\qquad [\R, \P] = \P.
\end{equation}

Infinitesimal deformations of $\g$ are classified by the second
Chevalley--Eilenberg cohomology group $H^2(\g;\g)$.  We are however
only interested in deformations which are kinematical; that is, such
that the brackets involving $R_{ab}$ are not deformed.  Infinitesimal
kinematical deformations are classified by the relative
Chevalley--Eilenberg cohomology group $H^2(\g,\s;\g)$.  Since
$\s \cong \so(D)$ is simple for $D\geq 5$, the Hochschild--Serre
factorisation theorem says that $H^2(\g;\g) \cong H^2(\g,\s;\g)$.
Therefore we see that all deformations of $\g$ are \emph{a priori}
kinematical.  A similar argument shows that this is also the case for
any $D \neq 2$.

Let $\h$ denote the abelian ideal generated by $B_a$, $P_a$ and
$H$. The cohomology group $H^2(\g,\s;\g)$ can be computed from the
$\s$-invariant complex
\begin{equation}
  C^\bullet := C^\bullet (\h;\g)^\s = \left\{\text{$\s$-equivariant linear maps }
    \Lambda^\bullet \h \to \g \right\} \cong \left(\Lambda^\bullet
    \h^* \otimes \g\right)^\s.
\end{equation}
Let $\beta_a$, $\pi_a$ and $\eta$ denote the canonical dual basis for
$\h^*$.  The differential $\d : C^p \to C^{p+1}$ is defined on
generators by
\begin{equation}
  \d B_a = \d P_a = \d H = \d \beta_a = \d \pi_a = \d \eta = 0
  \qquad\text{and}\qquad \d R_{ab} = \beta_a B_b - \beta_b B_a + \pi_a
  P_b - \pi_b P_a
\end{equation}
and extended to $C^\bullet$ as an odd derivation.  It is clear by
inspection that $\d^2 = 0$ on generators, and since it is an even
derivation, it is identically zero.

We proceed to enumerate the cochains in $C^p$ for $p\leq 3$.  We need
to calculate the action of the differential $\d : C^1 \to C^2$ and $\d
: C^2 \to C^3$ and in addition the Nijenhuis--Richardson brackets
$[\![-,-]\!]: C^2 \times C^2 \to C^3$.

\subsection{The Chevalley--Eilenberg cochains}
\label{sec:chev-eilenb-coch}

All $\so(D)$-invariant tensors are built out of $\delta_{ab}$ and
$\epsilon_{a_1\dots a_D}$.  We use the Einstein summation
convention in that repeated indices are summed over.  In the absence
of ambiguities, we also use an abbreviated notation for cochains
where we omit indices and assume that they are contracted with the
invariant tensors in the only way possible.

The $0$-cochains is the $\s$-invariant subset of $\g$, which is
one-dimensional and spanned by $H$.  Every $0$-cochain is a cocycle,
since $\d H = 0$.  There are no $1$-coboundaries.

The $1$-cochains are $\s$-equivariant linear maps $\h \to \g$.  A
basis for the $1$-cochains are given in Table~\ref{tab:basis-C1},
where we identify linear maps $\h\to \g$ with elements of $\h^*
\otimes \g$ and where $\eta H = \eta \otimes H$, $\beta B = \beta_a
\otimes B_a$, et cetera.  We see that all $1$-cochains are cocycles
and hence there are no $2$-coboundaries.

\begin{table}[h!]
  \centering
  \caption{Basis for $C^1(\h;\g)^{\s}$}
  \label{tab:basis-C1}
  \begin{tabular}{*{5}{>{$}c<{$}}}
    a_1& a_2 & a_3 & a_4 & a_5 \\\hline
    \eta H & \beta B & \beta P & \pi B & \pi P\\
  \end{tabular}
\end{table}

A basis for the $2$-cochains are given in Table~\ref{tab:basis-C2} as
elements of $\Lambda^2\h^*\otimes \g$, where $\beta\pi R = \beta_a
\wedge \pi_b \otimes R_{ab}$, et cetera.

\begin{table}[h!]
  \centering
  \caption{Basis for $C^2(\h;\g)^{\s}$}
  \label{tab:basis-C2}
  \begin{tabular}{*{8}{>{$}c<{$}}}
    c_1 & c_2 & c_3 & c_4 & c_5 & c_6 & c_7 & c_8 \\\hline
    \eta \beta B & \eta \beta P & \eta \pi B & \eta \pi P & \tfrac12 \beta\beta R & \beta\pi R & \tfrac12 \pi\pi R & \beta\pi H\\
  \end{tabular}
\end{table}

Finally, a basis for the $3$-cochains are given in
Table~\ref{tab:basis-C3} as elements of $\Lambda^3 \h^* \otimes \g$,
but in an abbreviated notation where, for example, $\beta\pi\beta B =
\beta_a \wedge \pi_a \wedge \beta_b \otimes B_b$, et cetera.  Those
cochains in the second row are only present when $D=5$ and are given
by $\epsilon \beta\beta\beta R = \epsilon_{abcde} \beta_a \wedge
\beta_b \wedge \beta_c \otimes R_{de}$, et cetera.  They turn out
to play no role in the calculations.

\begin{table}[h!]
  \centering
  \caption{Basis for $C^3(\h;\g)^{\s}$}
  \label{tab:basis-C3}
  \begin{tabular}{*{8}{>{$}c<{$}}}
    b_1 & b_2 & b_3 & b_4 & b_5 & b_6 & b_7 & b_8 \\\hline
   \eta\beta\pi H & \eta\beta\beta R & \eta\beta\pi R & \eta\pi\pi  R & \beta\pi\beta B & \beta\pi\beta P & \beta\pi\pi B & \beta\pi\pi P \\[10pt]
    & &  b'_9 &  b'_{10} &  b'_{11} &  b'_{12} & &  \\\hline
    & & \epsilon \beta\beta\beta R& \epsilon \beta\beta\pi R & \epsilon \beta\pi\pi R & \epsilon \pi\pi\pi R & &
  \end{tabular}
\end{table}

The last piece of data that we need is the Nijenhuis--Richardson bracket
$[\![-,-]\!]: C^2 \times C^2 \to C^3$, defined by
\begin{equation}
  [\![\lambda,\mu]\!] := \lambda \bullet \mu + \mu \bullet \lambda,
\end{equation}
where $\bullet$ is the operation defined on monomials by
\begin{equation}
  (\alpha \otimes X) \bullet (\beta \otimes Y) := (\alpha \wedge
  \iota_X \beta) \otimes Y,
\end{equation}
for $\alpha,\beta \in \Lambda^2\h^*$ and $X,Y \in \g$ and the
derivation $\iota_X$ denotes contraction with $X$.

\begin{table}[h!]
  \setlength{\extrarowheight}{4pt}
  \centering
  \caption{Nijenhuis--Richardson $\bullet$}
  \label{tab:NR-dot}
  \begin{tabular}{>{$}c<{$}|*{8}{>{$}c<{$}}}
    \bullet & c_1 & c_2 & c_3 & c_4 & c_5 & c_6 & c_7 & c_8 \\\hline
    c_1 & \zero & \zero & \zero & \zero & b_2 & b_3 & \zero & b_1 \\
    c_2 & \zero & \zero & \zero & \zero & \zero & b_2 & b_3 & \zero\\
    c_3 & \zero & \zero & \zero & \zero & b_3 & b_4 & \zero & \zero\\
    c_4 & \zero & \zero & \zero & \zero & \zero & b_3 & b_4 & b_1 \\
    c_5 & \zero & \zero & \zero & \zero & \zero & \zero & \zero & \zero\\
    c_6 & \zero & \zero & \zero & \zero & \zero & \zero & \zero & \zero \\
    c_7 & \zero & \zero & \zero & \zero & \zero & \zero & \zero & \zero \\
    c_8 & b_5 & b_6 & b_7 & b_8 & \zero & \zero & \zero & \zero \\
  \end{tabular}
\end{table}

Table~\ref{tab:NR-dot} collects the calculations of $c_i \bullet c_j$,
from where we can read off $[\![c_i, c_j]\!]$ by symmetrisation:
\begin{equation}
  \label{eq:NR-bracket}
  \begin{aligned}[m]
    [\![c_1, c_5]\!] &= b_2 \\
    [\![c_1, c_6]\!] &= b_3 \\
    [\![c_1, c_8]\!] &= b_1 + b_5 \\
    [\![c_2, c_6]\!] &= b_2\\    
  \end{aligned}
  \qquad\qquad
  \begin{aligned}[m]
    [\![c_2, c_7]\!] &= b_3 \\
    [\![c_2, c_8]\!] &= b_6 \\
    [\![c_3, c_5]\!] &= b_3 \\
    [\![c_3, c_6]\!] &= b_4\\    
  \end{aligned}
  \qquad\qquad
  \begin{aligned}[m]
    [\![c_3, c_8]\!] &= b_7 \\
    [\![c_4, c_6]\!] &= b_3 \\
    [\![c_4, c_7]\!] &= b_4 \\
    [\![c_4, c_8]\!] &= b_1 + b_8\\    
  \end{aligned}
\end{equation}

\subsection{Infinitesimal deformations}
\label{sec:infin-deform}

The action of the Chevalley--Eilenberg differential $\d: C^2 \to C^3$
is given by
\begin{equation}
  \label{eq:CE-d-2-3}
  \d c_1 = \d c_2 = \d c_3 = \d c_4 = \d c_8 = 0 \qquad \d c_5 = -b_6 \qquad \d
  c_6 = b_5 - b_8 \qquad\text{and}\qquad \d c_7 = b_7.
\end{equation}

Therefore the five-dimensional space of $2$-cocycles is spanned by
$c_1,c_2,c_3,c_4,c_8$. Since there are no $2$-coboundaries, this is
also the cohomology.  Therefore the most general infinitesimal
deformation is given by a linear combination
\begin{equation}
  \varphi_1 = t_1 c_1 + t_2 c_2 + t_3 c_3 + t_4 c_4 + t_5 c_8.
\end{equation}

\subsection{Obstructions}
\label{sec:obstructions}

The first obstruction to integrability is given by
\begin{equation}
  \tfrac12 [\![\varphi_1,\varphi_1]\!] = t_1 t_5 (b_1 + b_5) + t_2 t_5
  b_6 + t_3 t_5 b_7 + t_4 t_5 (b_1 + b_8).
\end{equation}
Projecting to $H^3$, we see from Equation~\eqref{eq:CE-d-2-3} that $[b_6]=[b_7]=0$ and that
$[b_5]=[b_8]$, so that
\begin{equation}
  \left[\tfrac12 [\![\varphi_1,\varphi_1]\!]\right] = (t_1+t_4) t_5
  ([b_1] + [b_5]),  
\end{equation}
so the obstruction vanishes if and only if
\begin{equation}
  \label{eq:quads}
  t_5 (t_1 + t_4) = 0.
\end{equation}
If that equation is satisfied, $\tfrac12 [\![\varphi_1,\varphi_1]\!] =
\d \varphi_2$, where
\begin{equation}
  \varphi_2 =t_1 t_5 c_6 - t_2 t_5 c_5 + t_3 t_5 c_7.
\end{equation}
The next obstruction is $[\![\varphi_1, \varphi_2]\!]$, which is seen
to vanish exactly provided that Equation~\eqref{eq:quads} is
satisfied.  This means that we can take $\varphi_3 = 0$.  The next
obstruction is $\tfrac12 [\![\varphi_2, \varphi_2]\!]$, which is seen
to vanish identically because $c_5, c_6, c_7$ have vanishing
Nijenhuis--Richardson brackets.  This means that the deformation
\begin{equation}
  \label{eq:varphi}
  \varphi = t_1 c_1 + t_2 c_2 + t_3 c_3 + t_4 c_4 - t_2 t_5 c_5 + t_1
  t_5 c_6  + t_3 t_5 c_7 + t_5 c_8
\end{equation}
defines a Lie algebra provided that Equation~\eqref{eq:quads} is
satisfied.  That equation has two branches, depending on whether or
not $t_5 = 0$.

\subsection{Deformations with $t_5= 0$}
\label{sec:deform-with-t_5=}

In this case, the deformation is
\begin{equation}
  \varphi = t_1 c_1 + t_2 c_2 + t_3 c_3 + t_4 c_4,
\end{equation}
which leads to the Lie brackets (in abbreviated notation):
\begin{equation}
  \begin{split}
    [H,\B] &= t_1 \B + t_2 \P\\
    [H,\P] &= t_3 \B + t_4 \P.
  \end{split}
\end{equation}
As in the $D=3$ case, we can bring these to normal forms depending on
the value of the discriminant $\delta := (t_1 - t_4)^2 + 4 t_2 t_3$:
\begin{enumerate}
\item $\delta > 0$ (or $\delta = 0$ and diagonalisable):
  \begin{equation}
    \label{eq:branch-1-1}
    \boxed{[H,\B] = \gamma \B \qquad\text{and}\qquad [H,\P] = \P,}
  \end{equation}
  where $\gamma \in [-1,1]$.  The case $\gamma = -1$ is the
  higher-dimensional version of the \textbf{(lorentzian) Newton Lie
  algebra}.

\item $\delta = 0$ (and not diagonalisable):
  \begin{equation}
    \label{eq:branch-1-2}
    \boxed{[H,\B] = \B + \P \qquad\text{and}\qquad [H,\P] = \P,}
  \end{equation}
  or
  \begin{equation}
    \label{eq:galilean}
    \boxed{[H,\B] = \P,}
  \end{equation}
  which is the \textbf{galilean algebra}.
  
\item $\delta < 0$:
  \begin{equation}
    \label{eq:branch-1-3}
    \boxed{[H,\B] = \alpha \B + \P \qquad\text{and}\qquad [H,\P] = -\B
      + \alpha \P,}
  \end{equation}
  where $\alpha \geq 0$.  The case $\alpha = 0$ is the
  higher-dimensional version of the \textbf{(euclidean) Newton Lie
  algebra}.
\end{enumerate}

\subsection{Deformations with $t_5 \neq 0$}
\label{sec:deform-with-t_5-neq-0}

In this case, Equation~\eqref{eq:quads} forces $t_1 + t_4 = 0$, so
that the deformation is
\begin{equation}
  \varphi = t_1 (c_1 - c_4) + t_2 c_2 + t_3 c_3 - t_2 t_5 c_5 + t_1
  t_5 c_6  + t_3 t_5 c_7 + t_5 c_8,
\end{equation}
which leads to the Lie brackets
\begin{equation}
  \begin{aligned}[m]
    [H, \B] &= t_1 \B + t_2 \P\\
    [H, \P] &= t_3 \B - t_1 \P
  \end{aligned}
  \qquad\qquad
  \begin{aligned}[m]
    [\B, \B] &= -t_2 t_5 \R\\
    [\P, \P] &=  t_3 t_5 \R
  \end{aligned}
  \qquad\qquad
    [\B, \P] = t_5 H + t_1 t_5 \R.
\end{equation}

In order to bring these Lie brackets to normal form, it proves
useful to study the action of those automorphisms of $\h$ which
commute with the action of $\s$.  This is similar to what happens in
$D=3$ and we refer to \cite{JMFKinematical3D} for a more detailed
description of the method.

The subgroup $G$ of automorphisms of $\h$ which commutes with the
$\s$-action is $\GL(\RR^2) \times \RR^\times$ acting on generators as
follows:
\begin{equation}
  (\B, \P, H) \mapsto (\B, \P, H)
  \begin{pmatrix}
    a & b & \zero \\ c & d & \zero \\ \zero & \zero & \lambda
  \end{pmatrix}
  \qquad\text{where}\quad
  \begin{pmatrix}
    a & b \\ c & d
  \end{pmatrix} \in \GL(\RR^2) \quad\text{and}\quad \lambda \in
  \RR^\times.
\end{equation}
The induced action on $\h^*$ is given by
\begin{equation}
  (\bbeta, \bpi, \eta) \mapsto (\bbeta, \bpi, \eta)
  \begin{pmatrix}
    d/\Delta & -c/\Delta & \zero \\ -b/\Delta & a/\Delta & \zero \\
    \zero & \zero & \lambda^{-1}
  \end{pmatrix}
  \qquad\text{where}\quad \Delta := ad - bc.
\end{equation}
From this one reads off how $G$ acts on $C^2$:
\begin{equation}
  \begin{split}
    c_1 + c_4 &\mapsto \tfrac1\lambda (c_1 + c_4)\\
    c_1 - c_4 &\mapsto \tfrac1{\lambda\Delta} \left( (a d + b c)
      (c_1 - c_4) + 2 c d c_2 - 2 a b c_3 \right)\\
    c_2 &\mapsto \tfrac1{\lambda\Delta} \left( bd (c_1 - c_4) + d^2 c_2 - b^2 c_3 \right)\\
    c_3 &\mapsto \tfrac1{\lambda\Delta} \left( -a c (c_1 - c_4) - c^2
      c_2 + a^2 c_3 \right)\\
    c_5 &\mapsto \tfrac1{\Delta^2} \left( d^2 c_5 - bd c_6 + b^2 c_7 \right)\\
    c_6 &\mapsto  \tfrac1{\Delta^2} \left( -2 c d c_5 + (a d + b c)
      c_6 - 2 a b c_7 \right)\\
    c_7 &\mapsto  \tfrac1{\Delta^2} \left( c^2 c_5 - ac c_6 + a^2 c_7 \right)\\
    c_8 &\mapsto \lambda \Delta c_8,\\
  \end{split}
\end{equation}
and from this we arrive at how $G$ acts on the deformation parameters
$(t_1,t_2,t_3,t_5)$:
\begin{equation}
  \begin{pmatrix}
    t_1 \\ t_2 \\ t_3 \\ t_5
  \end{pmatrix}
  \mapsto
  \frac1{\lambda\Delta}
  \begin{pmatrix}
    ad+bc & bd & -a c &  \zero \\
    2 c d & d^2 & -c^2 & \zero \\
    -2 a b & -b^2 & a^2 & \zero \\
    \zero & \zero & \zero & \lambda^2 \Delta^2
  \end{pmatrix}
  \begin{pmatrix}
    t_1 \\ t_2 \\ t_3 \\ t_5
  \end{pmatrix}.
\end{equation}
The representation $\rho$ of $\GL(\RR^2)$ defined by
\begin{equation}
  A =
  \begin{pmatrix}
    a & b \\ c & d 
  \end{pmatrix}
  \mapsto
  \frac1\Delta
  \begin{pmatrix}
    ad+bc & bd & -a c \\
    2 c d & d^2 & -c^2 \\
    -2 a b & -b^2 & a^2 \\
  \end{pmatrix}
\end{equation}
is not faithful -- having as kernel the scalar matrices -- and
preserves the lorentzian inner product
\begin{equation}
  K =
  \begin{pmatrix}
    2 & \zero & \zero \\
    \zero & \zero & 1\\
    \zero & 1 & \zero
  \end{pmatrix}.
\end{equation}
With a suitable choice of $\lambda$, we can use such three-dimensional
Lorentz transformations to bring $\t = (t_1, t_2, t_3)$ to one of the
following normal forms, each one labelling an orbit of $G$ on the
space of such parameters:
\begin{enumerate}
\item the \textbf{zero} orbit, where $\t = (0,0,0)$;
\item the \textbf{spacelike} orbit, where $\t = (1,0,0)$;
\item the \textbf{timelike} orbit, where $\t = (0,1,-1)$; and
\item the \textbf{lightlike} orbit, where $\t = (0,0,1)$.
\end{enumerate}
This still leaves the possibility to act with a (non-zero) scalar
matrix to set $t_5 = \pm 1$, since scalar matrices in $\GL(\RR^2)$
have positive determinant.  We discuss each one of these cases in
turn.

\subsubsection{The zero branch}
\label{sec:zero-branch}

In this case we can actually set $t_5 = 1$ without loss of generality.
The non-zero Lie brackets are
\begin{equation}
  \label{eq:carroll}
  \boxed{[\B,\P] = H,}
\end{equation}
which is the higher-dimensional analogue of the \textbf{Carroll
  algebra}.

\subsubsection{The spacelike branch}
\label{sec:spacelike-branch}

Here we can also set $t_5 = 1$ without loss of generality, and the
non-zero Lie brackets are
\begin{equation}
  [H, \B] = \B \qquad [H,\P] = -\P \qquad\text{and}\qquad  [\B, \P] = H + \R.
\end{equation}
This Lie algebra is isomorphic to $\so(D+1,1)$.  If we change basis so
that $(\B,\P) \mapsto (\frac1{\sqrt2} (\B+\P), \frac1{\sqrt2}
(\B-\P))$, then it takes the more standard form
\begin{equation}
  \label{eq:hyperbolic}
  \boxed{[H, \B] = \P \qquad [H,\P] = \B \qquad [\B,\B] = \R \qquad
    [\B, \P] = -H \qquad \text{and}\qquad [\P,\P] = -\R.}
\end{equation}

\subsubsection{The timelike branch}
\label{sec:timelike-branch}

In this case, the non-zero Lie brackets are, for $\varepsilon = \pm 1$,
\begin{equation}
  \label{eq:sphere}
  \boxed{
  \begin{aligned}[m]
    [H, \B] &= -\varepsilon \P\\
    [H, \P] &=  \varepsilon \B
  \end{aligned}
  \qquad\qquad
  \begin{aligned}[m]
    [\B, \B] &= \varepsilon \R\\
    [\P, \P] &= \varepsilon \R
  \end{aligned}
  \qquad\qquad
    [\B, \P] = H.}
\end{equation}
These Lie algebras are isomorphic to $\so(D+2)$ (for $\varepsilon=-1$)
or $\so(D,2)$ (for $\varepsilon=1$).

\subsubsection{The lightlike branch}
\label{sec:lightlike-branch}

In this case, the non-zero Lie brackets are, for $\varepsilon = \pm 1$,
\begin{equation}
  \label{eq:poincare}
  \boxed{
    [H, \P] = \varepsilon \B \qquad [\P, \P] =  \varepsilon \R  \qquad\text{and}\qquad [\B, \P] = H,}
\end{equation}
after redefining $H$.  These Lie algebras are isomorphic to the
euclidean Lie algebra $\e$ for $\varepsilon=-1$, or the Poincaré Lie
algebra $\p$ for $\varepsilon=1$.

\subsection{Invariant inner products}
\label{sec:invar-inner-prod}

We shall now investigate the existence of invariant inner products on
the kinematical Lie algebras determined in this section.  We remind
the reader that by an invariant inner product on a Lie algebra $\g$ we
mean a non-degenerate symmetric bilinear form
$(-,-): \g \times \g \to \RR$ which is ``associative''; that is,
\begin{equation}\label{eq:assoc}
  ([x,y],z) = (x,[y,z]) \qquad\text{for all~} x,y,z \in \g.
\end{equation}
The Killing form is associative, but by Cartan's semisimplicity
criterion, it is only non-degenerate for semisimple Lie algebras.  This
means that the simple Lie algebras $\so(D+1,1)$, $\so(D+2)$ and $\so(D,2)$
admit invariant inner products: namely, any non-zero multiple of the
Killing form.  It turns out that these are the only kinematical
Lie algebras which do.  To prove this, rather than appealing to any
general structural results, we simply exploit the associativity
condition \eqref{eq:assoc}.

We shall first of all show that no kinematical Lie algebra where $\B$
and $\P$ span an abelian ideal can admit an invariant inner product.
This rules out the first five rows in Table~\ref{tab:summary}.
Indeed, let $(-,-)$ be an associative symmetric bilinear form.  We
show that it is degenerate.  To this end, let $X,Y$ be any of
$B,P$ and consider (in abbreviated notation)
\begin{equation}
  (X, Y) = ([R,X],Y) = (R, [X,Y]) = 0,
\end{equation}
where we have used associativity and the fact that $X,Y$ are vectors
under rotations.  By rotational invariance, the only possible non-zero
inner products involving $B$ and $P$ are of the form $(B,B)$, $(B,P)$
and $(P,P)$, and as we have just seen, these are zero.  Therefore
$(-,-)$ is degenerate.

Any associative symmetric bilinear form in the Carroll algebra is
degenerate, since $(H,-) = 0$.  Indeed, by rotational invariance, the
only possible non-zero inner product of $H$ is with itself, but then
\begin{equation}
  \delta_{ab} (H,H) = (H, [B_a, P_b]) = ([H,B_a],P_b) = 0.
\end{equation}

It remains to consider the euclidean and Poincaré algebras.  So let
$(-,-)$ be an associative symmetric bilinear form on either $\e$ or
$\p$ and let us calculate $(H,H)$, which is the only possibly non-zero
rotationally invariant inner product involving $H$:
\begin{equation}
  \delta_{ab} (H,H) = ([B_a, P_b],H) = (B_a, [P_b,H]) = 0.
\end{equation}

\section{Kinematical Lie algebras with $D=4$}
\label{sec:kinem-lie-algebr}

In this section $\g$ denotes the static kinematical Lie algebra for
$D=4$. The case $D=4$ is slightly more complicated due to the fact
that the rotation subalgebra $\so(4)$ is not simple, but rather
$\so(4) \cong \so(3) \oplus \so(3)$. This is due to the existence of
the Hodge star, an $\so(4)$-invariant linear map
$\star : \Lambda^2\RR^4 \to \Lambda^2\RR^4$ which obeys
$\star^2 = \id$ and hence decomposes $\Lambda^2\RR^4$ into its
eigenspaces $\Lambda^2_\pm$, each one corresponding to an $\so(3)$
subalgebra.

Let
$R_{ab}^\pm := \tfrac12 \left( R_{ab} \pm \tfrac12 \epsilon_{abcd}
  R_{cd}\right)$ span $\s \cong \so(4)$ and $B_a,P_a,H$ span the
abelian ideal $\h$ of $\g$.  As usual we choose the canonical
dual basis $\beta_a, \pi_a, \eta$ for $\h^*$.

The non-zero Lie brackets in that basis are given by
\begin{equation}
  \begin{split}
    [R^\pm_{ab}, R^\pm_{cd}] &= [R_{ab}, R_{cd}]^\pm\\
    [R^\pm_{ab}, B_c] &= \tfrac12 \left(\delta_{bc} B_a - \delta_{ac}
      B_b \mp \epsilon_{abcd} B_d \right)\\
    [R^\pm_{ab}, P_c] &= \tfrac12 \left(\delta_{bc} P_a - \delta_{ac}
      P_b \mp \epsilon_{abcd} P_d \right)\\
  \end{split} 
\end{equation}

\subsection{The Chevalley--Eilenberg complex}
\label{sec:chev-eilenb-compl}

The fact that $\s$ is semisimple suffices for the Hochschild--Serre 
decomposition theorem and we may calculate the infinitesimal
deformations from the $\s$-invariant subcomplex $C^\bullet :=
C^\bullet(\h;\g)^\s$.  In particular this shows that all deformations
are automatically kinematical.

We now proceed to enumerate bases for the spaces of cochains, noting
that $C^0$ is spanned by $H$.  The dimensions of $C^1$, $C^2$ and
$C^3$ are 5, 11 and 19, respectively. Natural bases are tabulated
below.  In Table~\ref{tab:basis-C3p}, the cochains in the second row
involve the $\epsilon$ tensor, so that, for example, $\epsilon
\beta\beta \pi B = \epsilon_{abcd}\beta_a \wedge \beta_b \wedge \pi_c
\otimes B_d$, et cetera.

\begin{table}[h!]
  \centering
  \caption{Basis for $C^1(\h;\g)^{\s}$}
  \label{tab:basis-C1p}
  \begin{tabular}{*{5}{>{$}c<{$}}}
    a_1& a_2 & a_3 & a_4 & a_5 \\\hline
    \eta H & \beta B & \beta P & \pi B & \pi P\\
  \end{tabular}
\end{table}

\begin{table}[h!]
  \centering
  \caption{Basis for $C^2(\h;\g)^{\s}$}
  \label{tab:basis-C2p}
  \begin{tabular}{*{11}{>{$}c<{$}}}
    c_1 & c_2 & c_3 & c_4 & c_5 & c_6 & c_7 & c_8 & c_9 & c_{10} & c_{11} \\\hline
    \eta \beta B & \eta \beta P & \eta \pi B & \eta \pi P & \tfrac12 \beta\beta R^+ & \tfrac12 \beta\beta R^- & \beta\pi R^+ & \beta\pi R^- & \tfrac12 \pi\pi R^+ & \tfrac12 \pi\pi R^- & \beta\pi H\\
  \end{tabular}
\end{table}

\begin{table}[h!]
  \centering
  \caption{Basis for $C^3(\h;\g)^{\s}$}
  \label{tab:basis-C3p}
  \begin{tabular}{*{11}{>{$}c<{$}}}
    b_1 & b_2 & b_3 & b_4 & b_5 & b_6 & b_7 & b_8 & b_9 & b_{10} & b_{11} \\\hline
   \eta\beta\pi H & \eta\beta\beta R^+ & \eta\beta\beta R^-  & \eta\beta\pi R^+ & \eta\beta\pi R^- & \eta\pi\pi R^+ & \eta\pi\pi R^-& \beta\pi\beta B & \beta\pi\beta P & \beta\pi\pi B & \beta\pi\pi P \\[10pt]
    b_{12} &  b_{13} &  b_{14} &  b_{15} & b_{16} & b_{17} & b_{18} & b_{19} & & & \\\hline
    \epsilon \beta\beta\beta B & \epsilon \beta\beta\beta P & \epsilon \beta\beta\pi B & \epsilon \beta\beta\pi P & \epsilon \beta\pi\pi B & \epsilon \beta\pi\pi P & \epsilon \pi\pi\pi B & \epsilon \pi\pi\pi P & & & 
  \end{tabular}
\end{table}

The Chevalley--Eilenberg differential $\d : C^p \to C^{p+1}$ is
defined on generators by
\begin{equation}
  \d \eta = \d \beta_a = \d \pi_a = \d H = \d B_a = \d P_a = 0
\end{equation}
and
\begin{equation}
  \d R_{ab}^\pm = \tfrac12 \left(\beta_a B_b - \beta_b B_a + \pi_a P_b
    - \pi_b P_a \pm \epsilon_{abcd} (\beta_c B_d + \pi_c P_d) \right).
\end{equation}

In particular, it follows that $\d$ is identically zero on $C^0$ and
$C^1$, so that $B^2 = 0$.  On $C^2$ we find
\begin{equation}
  \label{eq:CE-C2}
  \begin{aligned}[m]
    \d c_5 &= -\tfrac12 b_9 + \tfrac14 (b_{12} + b_{15})\\
    \d c_6 &= -\tfrac12 b_9 - \tfrac14 (b_{12} + b_{15})\\
  \end{aligned}
  \qquad
  \begin{aligned}[m]
    \d c_7 &= \tfrac12 \left(b_8-b_{11} + b_{14} + b_{17}\right)\\
    \d c_8 &= \tfrac12 \left(b_8-b_{11} - b_{14} - b_{17}\right)\\
  \end{aligned}
  \qquad
  \begin{aligned}[m]
    \d c_9 &= \tfrac12 b_{10} + \tfrac14 (b_{16} + b_{19})\\
    \d c_{10} &= \tfrac12 b_{10} - \tfrac14 (b_{16} + b_{19}).\\
  \end{aligned}
\end{equation}
It follows from these expressions that $b_9 = \d(-c_5 - c_6)$ and
$b_{10}=\d(c_9 + c_{10})$ are coboundaries, as are $b_{16} + b_{19} =
\d (2(c_9-c_{10})$, $b_{12} + b_{15} = \d(2(c_5 - c_6)$, $b_8 - b_{11}
= \d(c_7 + c_8)$ and $b_{14} + b_{17} = \d(c_7 - c_9)$.  Indeed, these
coboundaries span $B^3$.

The last piece of data that we need is the Nijenhuis--Richardson
bracket $[\![-,-]\!]: C^2 \times C^2 \to C^3$.

\begin{table}[h!]
  \setlength{\extrarowheight}{4pt}
  \centering
  \caption{Nijenhuis--Richardson $\bullet$}
  \label{tab:NR-dot-4}
  \begin{tabular}{>{$}c<{$}|*{11}{>{$}c<{$}}}
    \bullet & c_1 & c_2 & c_3 & c_4 & c_5 & c_6 & c_7 & c_8 & c_9 & c_{10} & c_{11} \\\hline
    c_1 & \zero & \zero & \zero & \zero & b_2 & b_3 & b_4 & b_5 & \zero & \zero & b_1 \\
    c_2 & \zero & \zero & \zero & \zero & \zero & \zero & b_2 & b_3 & b_4 & b_5 & \zero\\
    c_3 & \zero & \zero & \zero & \zero & b_4 & b_5 & b_6 & b_7 & \zero & \zero & \zero\\
    c_4 & \zero & \zero & \zero & \zero & \zero & \zero & b_4 & b_5 & b_6 & b_7 & b_1 \\
    c_5 & \zero & \zero & \zero & \zero & \zero & \zero & \zero & \zero & \zero & \zero & \zero\\
    c_6 & \zero & \zero & \zero & \zero & \zero & \zero & \zero & \zero & \zero & \zero & \zero\\
    c_7 & \zero & \zero & \zero & \zero & \zero & \zero & \zero & \zero & \zero & \zero & \zero\\
    c_8 & \zero & \zero & \zero & \zero & \zero & \zero & \zero & \zero & \zero & \zero & \zero\\
    c_9 & \zero & \zero & \zero & \zero & \zero & \zero & \zero & \zero & \zero & \zero & \zero\\
    c_{10} & \zero & \zero & \zero & \zero & \zero & \zero & \zero & \zero & \zero & \zero & \zero\\
    c_{11} & b_8 & b_9 & b_{10} & b_{11} & \zero & \zero & \zero & \zero & \zero & \zero & \zero\\
  \end{tabular}
\end{table}

Table~\ref{tab:NR-dot-4} collects the calculations of
$c_i \bullet c_j$, from where we can read off $[\![c_i, c_j]\!]$ by
symmetrisation:
\begin{equation}
  \label{eq:NR-bracket-4}
  \begin{aligned}[m]
    [\![c_1, c_5]\!] &= b_2 \\
    [\![c_1, c_6]\!] &= b_3 \\
    [\![c_1, c_7]\!] &= b_4 \\
    [\![c_1, c_8]\!] &= b_5 \\
    [\![c_1, c_{11}]\!] &= b_1 + b_8\\    
  \end{aligned}
  \qquad\qquad
  \begin{aligned}[m]
    [\![c_2, c_7]\!] &= b_2 \\
    [\![c_2, c_8]\!] &= b_3 \\
    [\![c_2, c_9]\!] &= b_4 \\
    [\![c_2, c_{10}]\!] &= b_5\\
    [\![c_2, c_{11}]\!] &= b_9\\
  \end{aligned}
  \qquad\qquad
  \begin{aligned}[m]
    [\![c_3, c_5]\!] &= b_4 \\
    [\![c_3, c_6]\!] &= b_5 \\
    [\![c_3, c_7]\!] &= b_6 \\
    [\![c_3, c_8]\!] &= b_7\\    
    [\![c_3, c_{11}]\!] &= b_{10}\\    
  \end{aligned}
  \qquad\qquad
  \begin{aligned}[m]
    [\![c_4, c_7]\!] &= b_4 \\
    [\![c_4, c_8]\!] &= b_5 \\
    [\![c_4, c_9]\!] &= b_6 \\
    [\![c_4, c_{10}]\!] &= b_7\\
    [\![c_4, c_{11}]\!] &= b_1 + b_{11}.\\
  \end{aligned}
\end{equation}

\subsection{Infinitesimal deformations}
\label{sec:infin-deform-1}

From the action of the Chevalley--Eilenberg differential $\d$ on
$C^2$, described in Equation~\eqref{eq:CE-C2}, we find that
\begin{equation}
  H^2 \cong Z^2 =  \RR \left<c_1, c_2, c_3, c_4, c_{11}\right>.
\end{equation}
Therefore the most general infinitesimal deformation is
\begin{equation}
  \varphi_1 = t_1 c_1 + t_2 c_2 + t_3 c_3 + t_4 c_4 + t_5 c_{11}.
\end{equation}

\subsection{Obstructions}
\label{sec:obstructions-1}

The first obstruction is the class in $H^3$ of $\tfrac12
[\![\varphi_1, \varphi_1]\!]$, which we can calculate from the
explicit expressions \eqref{eq:NR-bracket-4} for the
Nijenhuis--Richardson bracket.  Doing so, we find
\begin{equation}
  \tfrac12 [\![\varphi_1, \varphi_1]\!] = t_1 t_5 (b_1 + b_8) + t_2
  t_5 b_9 + t_3 t_5 b_{10} + t_4 t_5 (b_1 + b_{11}).
\end{equation}
From Equation~\eqref{eq:CE-C2} we learn that
\begin{equation}
  B^3 = \RR\left<b_8-b_{11}, b_9, b_{10}, b_{12}+b_{15}, b_{14} +
    b_{17}, b_{16}+b_{19}\right>,
\end{equation}
so that in cohomology, $[b_8]=[b_{11}]$, $[b_9] = [b_{10}] = 0$,
$[b_{12}]=-[b_{15}]$, $[b_{14}] = - [b_{17}]$ and $[b_{16}] = -
[b_{19}]$.  Therefore in $H^3$,
\begin{equation}
  \left[\tfrac12 [\![\varphi_1, \varphi_1]\!] \right] = (t_1 + t_4)
  t_5 ([b_1] + [b_8]),  
\end{equation}
so that the obstruction vanishes if and only if
\begin{equation}
  \label{eq:quads-4}
  (t_1 + t_4) t_5 = 0.
\end{equation}
If this equation is satisfied,
\begin{equation}
 \tfrac12[\![\varphi_1,\varphi_1]\!] =  t_1 t_5 (b_8 - b_{11}) + t_2
 t_5 b_9 + t_3 t_5 b_{10},
\end{equation}
or, in other words, $\tfrac12[\![\varphi_1,\varphi_1]\!] = \d\varphi_2$ for
\begin{equation}
  \varphi_2 =t_1 t_5 (c_7 + c_8) - t_2 t_5 (c_5 + c_6) + t_3 t_5 (c_9
  + c_{10}).
\end{equation}
It follows from \eqref{eq:quads-4} that the next obstruction
$[\![\varphi_1,\varphi_2]\!]$ vanishes identically, so we can take
$\varphi_3=0$.  The next obstruction after that is $\tfrac12
[\![\varphi_2,\varphi_2]\!]$, but this also vanishes identically, so
that $\varphi_4 = 0$ and hence there are no further obstructions.  In
summary, the most general deformation is given by
\begin{equation}
  \label{eq:varphi-4}
  \varphi = t_1 c_1 + t_2 c_2 + t_3 c_3 + t_4 c_4 + t_5 c_{11} + t_1
  t_5 (c_7 + c_8) - t_2 t_5 (c_5 + c_6) + t_3 t_5 (c_9 + c_{10}),
\end{equation}
subject to Equation~\eqref{eq:quads-4}.

Notice that since $R^+ + R^- = R$, the sums of cochains appearing in
$\varphi$ are
\begin{equation}
    c_5 + c_6 = \tfrac12 \beta\beta R \qquad
    c_7 + c_8 = \beta \pi R \qquad\text{and}\qquad
    c_9 + c_{10} = \tfrac12 \pi\pi R,
\end{equation}
which means that the expression~\eqref{eq:varphi-4} above for
$\varphi$ coincides \emph{mutatis mutandis} with the one in
Equation~\eqref{eq:varphi}.  Since the conditions \eqref{eq:quads-4}
and \eqref{eq:quads} are identical, the rest of the analysis proceeds
as in the case $D\geq 5$.  As in $D\geq 5$, only the simple
kinematical Lie algebras admit an associative inner product.

\section{Deformations of the centrally extended static kinematical Lie
  algebra with $D\geq 5$}
\label{sec:deform-centr-extend}

Let $D\geq 5$ and let us consider the static kinematical Lie algebra
$\g$ defined (in abbreviated form) by Equation~\eqref{eq:static}.  We
shall show that it has a central extension with bracket
\begin{equation}
  [B_a, P_b] = \delta_{ab} Z\qquad (\text{or in abbreviated form }
  [\B,\P] = Z.)
\end{equation}
We now go on to classify the deformations of the centrally extended Lie
algebra $\tg = \g \oplus \RR Z$.

\subsection{Central extension of the static kinematical Lie algebra}
\label{sec:centr-extens-stat}

Central extensions of $\g$ are classified by $H^2(\g;\RR)$, which by
the Hochschild--Serre factorisation theorem is isomorphic to
$H^2(\h;\RR)^\s$ and this in turn can be computed from the subcomplex
$C^\bullet$ of $\s$-invariant cochains in $\Lambda^\bullet \h^*$.  The
first three spaces of cochains in the subcomplex are
\begin{equation}
  C^1 = \RR\left<\eta\right> \qquad C^2 = \RR\left<\beta\pi\right>
  \qquad\text{and}\qquad C^3 = \RR\left<\eta\beta\pi\right>,
\end{equation}
where we again use the abbreviated notation $\beta\pi = \beta_a \wedge
\pi_a$, et cetera.  Since $\h$ is abelian, the differential is
identically zero, so that $H^2 = C^2$ with cocycle representative
$\beta\pi$.  The universal central extension $\tg$ of $\g$ is thus
spanned by $R_{ab}, B_a, P_a, H, Z$ with non-zero brackets
\begin{equation}
  \label{eq:cex-static}
\boxed{[\R,\R] = \R \qquad [\R,\B] = \B \qquad [\R,\P] =\P
  \qquad\text{and}\qquad [\B,\P] = Z.}
\end{equation}

\subsection{The deformation complex}
\label{sec:deformation-complex}

Let $\th$ denote the ideal of $\tg$ spanned by $B_a, P_a, H, Z$ and
let $\s \cong \so(D)$ again be the rotational subalgebra.  By
Hochschild--Serre, the deformation complex can be taken to be the
$\s$-invariant subcomplex $C^\bullet = C^\bullet(\th;\tg)^\s$.  In
this section we describe this complex in a way useful for
calculations.  Let $\beta_a,\pi_a,\eta,\zeta$ be the canonical dual
basis for $\th^*$.

The Chevalley--Eilenberg differential is defined on generators as
follows:
\begin{equation}
  \label{eq:CE-diff}
  \begin{aligned}[m]
    \d R_{ab} &= \beta_a B_b - \beta_b B_a + \pi_a P_b - \pi_b P_a\\
    \d B_a &= - \pi_a Z\\
    \d P_a &= \beta_a Z\\
    \d Z &= \d H = 0\\
  \end{aligned}
  \qquad\qquad
  \begin{aligned}[m]
    \d \zeta &= -\beta_a \pi_a\\
    \d \beta_a &= 0\\
    \d \pi_a &= 0\\
    \d \eta &= 0.\\
  \end{aligned}
\end{equation}

Let $G \cong \GL(\RR^2) \ltimes \Aff(\Lambda^2\RR^2)$ denote the
subgroup of automorphisms of $\th$ which commutes with the action of
$\s$.  It leaves $\R$ invariant and acts on $\th$ as follows:
\begin{equation}
  (\B, \P, H, Z) \mapsto (\B, \P, H, Z)
  \begin{pmatrix}
    a & b & \zero & \zero \\ c & d & \zero & \zero \\ \zero & \zero &
    \lambda & \zero \\ \zero & \zero & \mu\Delta & \Delta
  \end{pmatrix},
\end{equation}
where
\begin{equation}
  \begin{pmatrix}
    a & b \\ c & d
  \end{pmatrix} \in \GL(\RR^2),\quad \Delta := a d - b c, \quad \mu \in \RR \quad\text{and}\quad \lambda \in
  \RR^\times.
\end{equation}
The induced action on $\th^*$ is given by
\begin{equation}
  (\bbeta, \bpi, \eta, \zeta) \mapsto (\bbeta, \bpi, \eta,\zeta)
  \begin{pmatrix}
    d/\Delta & -c/\Delta & \zero & \zero \\ -b/\Delta & a/\Delta &
    \zero & \zero \\
    \zero & \zero & \lambda^{-1} & -\lambda^{-1} \mu\\
    \zero & \zero & \zero & \Delta^{-1}
  \end{pmatrix}.
\end{equation}

We proceed to enumerate the cochains.  $C^0$ is spanned by $H,Z$.
The following tables enumerate the cochains in $C^1$, $C^2$ and
$C^3$.  The primed cochains in Table~\ref{tab:basis-CE-C3} are only
present for $D=5$ and as in the case of the static kinematical Lie
algebra they turn out not play any role in the calculations.

\begin{table}[h!]
  \centering
  \caption{Basis for $C^1(\th;\tg)^{\s}$}
  \label{tab:basis-CE-C1}
  \begin{tabular}{*{8}{>{$}c<{$}}}
    a_1& a_2 & a_3 & a_4 & a_5 & a_6 & a_7 & a_8 \\\hline
    \eta H & \eta Z & \zeta H & \zeta Z & \beta B & \beta P & \pi B & \pi P\\
  \end{tabular}
\end{table}

\begin{table}[h!]
  \centering
  \caption{Basis for $C^2(\th;\tg)^{\s}$}
  \label{tab:basis-CE-C2}
  \begin{tabular}{*{15}{>{$}c<{$}}}
    c_1 & c_2 & c_3 & c_4 & c_5 & c_6 & c_7 & c_8 & c_9 & c_{10} & c_{11} & c_{12} & c_{13} & c_{14} & c_{15}\\\hline 
    \eta\zeta H & \eta\zeta Z & \eta \beta B & \eta \beta P & \eta \pi B & \eta \pi P & \zeta \beta B & \zeta \beta P & \zeta \pi B & \zeta \pi P & \beta \pi H & \beta \pi Z & \tfrac12 \beta\beta R & \beta\pi R & \tfrac12 \pi\pi R \\
  \end{tabular}
\end{table}

\begin{table}[h!]
  \centering
  \caption{Basis for $C^3(\th;\tg)^{\s}$}
  \label{tab:basis-CE-C3}
  \begin{tabular}{*{11}{>{$}c<{$}}}
    b_1 & b_2 & b_3 & b_4 & b_5 & b_6 & b_7 & b_8 & b_9 & b_{10} & b_{11}\\\hline
    \eta\zeta\beta B & \eta\zeta\beta P & \eta\zeta\pi B & \eta\zeta\pi P & \eta\beta\pi H & \eta\beta\pi Z & \eta\beta\beta R & \eta\beta\pi R & \eta\pi\pi R & \zeta\beta\pi H & \zeta\beta\pi Z\\[10pt]
    b_{12} & b_{13} & b_{14} & b_{15} & b_{16} & b_{17} & b_{18} &  b'_{19} &  b'_{20} &  b'_{21} &  b'_{22} \\\hline
    \zeta\beta\beta R & \zeta\beta\pi R & \zeta\pi\pi R &\beta\pi\beta B & \beta\pi\beta P & \beta\pi\pi B & \beta\pi\pi P & \epsilon\beta\beta\beta R & \epsilon\beta\beta\pi R & \epsilon\beta\pi\pi R & \epsilon\pi\pi\pi R\\
  \end{tabular}
\end{table}

The Chevalley--Eilenberg differential is identically zero on $C^0$, so
that $B^1 = 0$.  The differential $\d : C^1 \to C^2$ is given on the
basis by
\begin{equation}
  \d a_1 = \d a_2 = \d a_6 = \d a_7 = 0 \qquad \d a_3 = - c_{11}
  \qquad \d a_4 = - c_{12} \qquad \d a_5 = c_{12} \qquad \d a_8 = c_{12},
\end{equation}
from where we see that $B^2 = \RR\left<c_{11},c_{12}\right>$.  The
differential $\d : C^2 \to C^3$ is given on the basis by
\begin{equation}
  \begin{aligned}[m]
    \d c_1 &= b_5 \\
    \d c_2 &= b_6 \\
    \d c_3 &= -b_6
  \end{aligned}
  \qquad
  \begin{aligned}[m]
    \d c_4 &= 0 \\
    \d c_5 &= 0 \\
    \d c_6 &= -b_6
  \end{aligned}
  \qquad
  \begin{aligned}[m]
    \d c_7 &= -b_{11} - b_{15} \\
    \d c_8 &= -b_{16} \\
    \d c_9 &= -b_{17}
  \end{aligned}
  \qquad
  \begin{aligned}[m]
    \d c_{10} &= -b_{11} - b_{18} \\
    \d c_{11} &= 0 \\
    \d c_{12} &= 0
  \end{aligned}
  \qquad
  \begin{aligned}[m]
    \d c_{13} &= -b_{16} \\
    \d c_{14} &= b_{15}-b_{18} \\
    \d c_{15} &= b_{17},
  \end{aligned}
\end{equation}
from where we see that $B^3 = \RR\left<b_5, b_6, b_{16}, b_{17},
  b_{11} + b_{15}, b_{11} + b_{18}\right>$ and that
\begin{equation}
  Z^2 = B^2 \oplus \H^2 \qquad\text{where}\qquad \H^2 := \RR \left<c_2 + c_3, c_2 + c_6, c_4, c_5, c_7 -
    c_{10} + c_{14}, c_8 - c_{13}, c_9 + c_{15}\right>.
\end{equation}
The subspace $\H^2$ is isomorphic to the cohomology, but we would
like to choose representative cocycles adapted to the action of $G$.
The action of $G$ on the complex can be read off from the action on
the generators and one finds that a convenient description of $\H^2$
is the following:
\begin{equation}
  \H^2 = \RR\left<2c_2 + c_3 + c_6\right> \oplus \RR \left<c_7 -
    c_{10} + c_{14}, c_8 - c_{13}, c_9 + c_{15}, c_3 - c_6, c_4, c_5 \right>,
\end{equation}
where, if we denote the above basis for $\H^2$ by
\begin{equation}
  (\be_1,\dots,\be_7) := (2c_2 + c_3 + c_6, c_7 -  c_{10} + c_{14}, c_8
  - c_{13}, c_9 + c_{15}, c_3 - c_6, c_4, c_5),
\end{equation}
then under the action of $G$,
\begin{equation}
  (\be_1, \dots, \be_7) = (\be_1, \dots, \be_7)
  \begin{pmatrix}
    \lambda^{-1} & \zero & \zero \\
    \zero & M_A & \zero \\
    \zero & -\lambda^{-1}\mu\Delta  M_A & \lambda^{-1}\Delta M_A
  \end{pmatrix},
\end{equation}
where
\begin{equation}
  \label{eq:MA}
  A =
  \begin{pmatrix}
    a & b \\ c & d
  \end{pmatrix} \qquad\text{and}\qquad
  M_A = \frac1{\Delta^2}
  \begin{pmatrix}
    ad + bc & bd & -ac\\
    2cd & d^2 & -c^2\\
    -2ab & -b^2 & a^2
  \end{pmatrix}.
\end{equation}
As we saw above the representation $A \mapsto M_A$ of $\GL(\RR^2)$ is
not faithful and has kernel the scalar matrices in $\GL(\RR^2)$, and
it preserves the lorentzian inner product with matrix
\begin{equation}
  \label{eq:lor-k}
  K =
  \begin{pmatrix}
    2 & \zero & \zero\\
    \zero & \zero & 1\\
    \zero & 1 & \zero
  \end{pmatrix}.
\end{equation}

The last piece of data that we shall need is the Nijenhuis--Richardson
bracket $[\![-,-]\!] : C^2 \times C^2 \to C^3$, which can be obtained
by symmetrisation from the $\bullet$ product tabulated in
Table~\ref{tab:NR-dot-CE}.

\begin{table}[h!]
  \setlength{\extrarowheight}{4pt}
  \centering
  \caption{Nijenhuis--Richardson $\bullet$}
  \label{tab:NR-dot-CE}
  \begin{tabular}{>{$}c<{$}|*{15}{>{$}c<{$}}}
    \bullet & c_1 & c_2 & c_3 & c_4 & c_5 & c_6 & c_7 & c_8 & c_9 & c_{10} & c_{11} & c_{12} & c_{13} & c_{14} & c_{15} \\\hline
    c_1 & \zero & \zero & b_{1} & b_{2} & b_{3} & b_{4} & \zero & \zero & \zero & \zero & \zero & \zero & \zero & \zero & \zero\\
    c_2 &\zero & \zero & \zero & \zero & \zero & \zero & b_{1} & b_{2} & b_{3} & b_{4} & \zero & \zero & \zero & \zero & \zero\\
    c_3 & \zero & \zero & \zero & \zero & \zero & \zero & b_{1} & b_{2} & \zero & \zero & b_{5} & b_{6} & b_{7} & b_{8} & \zero\\
    c_4 & \zero & \zero & \zero & \zero & \zero & \zero & \zero & \zero & b_{1} & b_{2} & \zero & \zero & \zero & b_{7} & b_{8}\\
    c_5 & \zero & \zero & \zero & \zero & \zero & \zero & b_{3} & b_{4} & \zero & \zero & \zero & \zero & b_{8} & b_{9} & \zero\\
    c_6 & \zero & \zero & \zero & \zero & \zero & \zero & \zero & \zero & b_{3} & b_{4} & b_{5} & b_{6} & \zero & b_{8} & b_{9}\\
    c_7 & \zero & \zero & -b_{1} & -b_{2} & \zero & \zero & \zero & \zero & \zero & \zero & b_{5} & b_{6} & b_{12} & b_{13} & \zero\\
    c_8 & \zero & \zero & \zero & \zero & -b_{1} & -b_{2} & \zero & \zero & \zero & \zero & \zero & \zero & \zero & b_{12} & b_{13}\\
    c_9 & \zero & \zero & -b_{3} & -b_{4} & \zero & \zero & \zero & \zero & \zero & \zero & \zero & \zero & b_{13} & b_{14} & \zero\\
    c_{10} & \zero & \zero & \zero & \zero & -b_{3} & -b_{4} & \zero & \zero & \zero & \zero & b_{10} & b_{11} & \zero & b_{13} & b_{14}\\
    c_{11} & b_{10} & b_{11} & b_{15} & b_{16} & b_{17} & b_{18} & \zero & \zero & \zero & \zero & \zero & \zero & \zero & \zero & \zero\\
    c_{12} &-b_{5} & -b_{6} & \zero & \zero & \zero & \zero & b_{15} & b_{16} & b_{17} & b_{18} & \zero & \zero & \zero & \zero & \zero\\
    c_{13} &\zero & \zero & \zero & \zero & \zero & \zero & \zero & \zero & \zero & \zero & \zero & \zero & \zero & \zero & \zero\\
    c_{14} &\zero & \zero & \zero & \zero & \zero & \zero & \zero & \zero & \zero & \zero & \zero & \zero & \zero & \zero & \zero\\
    c_{15} &\zero & \zero & \zero & \zero & \zero & \zero & \zero & \zero & \zero & \zero & \zero & \zero & \zero & \zero & \zero\\
  \end{tabular}
\end{table}

\subsection{Infinitesimal deformations and obstructions}
\label{sec:infin-deform-obstr}

We parametrise the $7$-dimensional space $\H^2$ of infinitesimal
deformations as
\begin{equation}
  \label{eq:inf-def-ce}
  \varphi_1 = t_1 (2c_2 + c_3 + c_6) + t_2 (c_7 - c_{10} + c_{14}) +
  t_3 (c_8 - c_{13}) + t_4 (c_9 + c_{15}) + t_5 (c_3 - c_6) + t_6 c_4
  + t_7  c_5.
\end{equation}
The first obstruction to integrability is $\tfrac12
[\![\varphi_1,\varphi_1]\!]$ whose vanishing (in cohomology, but in
this case also on the nose) is equivalent to the following system of
quadrics:
\begin{equation}
  \label{eq:quadrics-ce}
  \begin{split}
    0&= 2 t_1 t_2 - t_3 t_7 + t_4 t_6 \\
    0&= t_1 t_3 + t_3 t_5 - t_2 t_6\\
    0&= t_1 t_4 - t_4 t_5 + t_2 t_7.
  \end{split}
\end{equation}
Since when these equations are satisfied $[\![\varphi_1,\varphi_1]\!]
= 0$ (not just in cohomology) we can take $\varphi_2 = 0$ and
therefore there are no further obstructions to integrability.

To analyse this system of quadrics further we exploit the action
of the automorphism group $G$.  It follows from the action of $G$ on
$C^2$ that we can bring the triplet $\t = (t_5, t_6, t_7)$ to one of
four canonical forms, corresponding the causal type of $\t$ relative
to the lorentzian inner product defined by $K$ in \eqref{eq:lor-k}:
\begin{enumerate}
\item \textbf{Zero orbit}: $\t = (0,0,0)$
\item \textbf{Spacelike orbit}: $\t = (1,0,0)$
\item \textbf{Lightlike orbit}: $\t = (0,0,1)$
\item \textbf{Timelike orbit}: $\t = (0,1,-1)$
\end{enumerate}

\subsubsection{Zero orbit branch}
\label{sec:zero-orbit-branch}

Here $t_5= t_6 = t_7 = 0$ and the system \eqref{eq:quadrics-ce} of
quadrics becomes $t_1 t_2 = t_1 t_3 = t_1 t_4 = 0$, so we have two
cases to consider depending on whether or not $t_1 = 0$.

If $t_1 \neq 0$, then $t_2 = t_3 = t_4 = 0$ and by a judicious choice
of $\lambda \in \RR^\times$ we can set $t_1 = 1$ and the deformation
is
\begin{equation}
  \varphi = 2c_2 + c_3 + c_6
\end{equation}
which translates into the following additional brackets:
\begin{equation}
  \label{eq:branch-0-1}
  \boxed{[H,\B] = \B \qquad [H,\P] = \P \qquad\text{and}\qquad [H,Z] = 2 Z,}
\end{equation}
which we recognise as a non-central extension of the kinematical Lie
algebra \eqref{eq:branch-1-1} with $\gamma=1$.  Indeed, $Z$ spans an
ideal and quotienting by this ideal recovers the Lie algebra
\eqref{eq:branch-1-1} with $\gamma=1$.

If $t_1 = 0$, then we can use the automorphisms to bring
$\t = (t_2, t_3, t_4)$ to one of several canonical normal forms.
First of all, notice that on the three-dimensional subspace of such
$\t$ the group $G$ acts via $\t \mapsto M_A \t$, with $M_A$ the matrix
in \eqref{eq:MA}.  This defines an action of
$\GL(\RR^2)/\RR^\times \cong \SO_o(2,1)$, the identity component of
the three-dimensional Lorentz group.  Under the action of proper,
orthochronous Lorentz transformations, $\RR^3$ breaks up into the
following orbits:
\begin{enumerate}
\item \textbf{Zero orbit}: $\t= (0,0,0)$.  This corresponds to no
  deformation at all.
\item \textbf{Spacelike orbits}: $\t = (x,0,0)$ with $x>0$.  In this
  case, the deformation is given by
  \begin{equation}
    \varphi = x (c_7 - c_{10} + c_{14}) = x (\zeta \beta B - \zeta \pi
    P + \beta \pi R),
  \end{equation}
  so that the brackets are
  \begin{equation}
    [Z,\B] = x \B \qquad [Z,\P] = -x \P \qquad\text{and}\qquad [\B,\P]
    = Z + x \R.     
  \end{equation}
  We can rescale $Z \mapsto x^{-1} Z$ and $\P \mapsto x^{-1} \P$ and
  in this way set $x = 1$.  The resulting Lie brackets are
  \begin{equation}
    \label{eq:0-spacelike}
    \boxed{[Z,\B] = \B \qquad [Z,\P] = - \P \qquad\text{and}\qquad
      [\B,\P] = Z + \R,}
  \end{equation}
  which is isomorphic to a trivial central extension of $\so(D+1,1)$
  with central element $H$.

\item \textbf{Lightlike branches}: $\t = (0,0,\varepsilon)$, where
  $\varepsilon = \pm 1$.  The deformation cochain is
  \begin{equation}
    \varphi = \varepsilon ( c_9 + c_{15} ) = \varepsilon ( \zeta \pi B +
    \tfrac12 \pi\pi R),
  \end{equation}
  with Lie brackets
  \begin{equation}
    \label{eq:0-lightlike}
    \boxed{[Z,\P] = \varepsilon \B \qquad [\P,\P] = \varepsilon \R
      \qquad\text{and}\qquad [\B,\P] = Z.}
  \end{equation}
  We recognise these algebras as trivial central extensions of the
  euclidean (for $\varepsilon=-1$) or Poincaré (for $\varepsilon=+1$)
  algebras, with central element $H$.

\item \textbf{Timelike branches}: $\t = (0,x,-x)$, where $x \in
  \RR^\times$.  The deformation cochain is
  \begin{equation}
    \varphi = x (c_8 - c_9 - c_{13} - c_{15} ) = x ( \zeta \beta P -
    \zeta \pi B - \tfrac12 \beta\beta R - \tfrac12 \pi\pi R ),
  \end{equation}
  and Lie brackets
  \begin{equation}
    [Z,\B] = x \P \qquad [Z,\P] = - x \B \qquad [\B,\B] = -x \R
    \qquad\text{and}\qquad [\P,\P] = - x \R.
  \end{equation}
  Let $\varepsilon = -x/|x|$ be (minus) the sign of $x$.  Rescaling $\B \mapsto
  |x|^{-1/2} \B$, $\P \mapsto |x|^{-1/2} \P$ and $Z \mapsto |x|^{-1}
  Z$, we may bring the brackets to one of two forms, depending on
  $\varepsilon$:
  \begin{equation}
    \label{eq:0-timelike}
    \boxed{[Z,\B] = -\varepsilon \P \qquad [Z,\P] = \varepsilon \B \qquad [\B,\B] =
      \varepsilon \R \qquad [\P,\P] = \varepsilon \R
      \qquad\text{and}\qquad  [\B,\P] = Z.}
  \end{equation}
  We recognise these Lie algebras as trivial central extensions of
  $\so(D+2)$ (for $\varepsilon=-1$) or $\so(D,2)$ (for
  $\varepsilon=1$) with central element $H$.
\end{enumerate}

\subsubsection{Spacelike orbit branches}
\label{sec:spac-orbit-branch}

Here $t_5 = 1$ and $t_6 = t_7 = 0$ and the system
of quadrics \eqref{eq:quadrics-ce} becomes
\begin{equation}
  (t_1+1) t_3 = 0 \qquad (t_1 -1) t_4 = 0 \qquad\text{and}\qquad t_1
  t_2 = 0.
\end{equation}
We therefore have several branches depending on the value of $t_1$.
\begin{enumerate}
\item If $t_1 \neq 0,\pm 1$, then $t_2 = t_3 = t_4 = 0$ and the
  deformation is
  \begin{equation}
    \varphi = 2 t_1 c_2 + (t_1 + 1) c_3 + (t_1 -1) c_6 = 2 t_1 \eta
    \zeta Z + (t_1 + 1) \eta\beta B + (t_1 -1) \eta\pi P,
  \end{equation}
  with brackets
  \begin{equation}
    [H,\B] = (t_1 + 1) \B \qquad [H,\P] = (t_1 -1) \P
    \qquad\text{and}\qquad [H,Z] = 2 t_1 Z.
  \end{equation}
  We can bring this to a normal form by rescaling $H$ and, if
  necessary, interchanging $\B$ and $\P$ and changing the sign of $Z$:
  \begin{equation}
    \label{eq:spacelike-1}
    \boxed{[H,\B] = \gamma\B \qquad [H,\P] = \P \qquad [H,Z] = (1 +
      \gamma) Z \qquad\text{and}\qquad [\B,\P] = Z \qquad \gamma \in (-1,1).}
  \end{equation}
  This Lie algebra is isomorphic to a non-central extension of the Lie
  algebra \eqref{eq:branch-1-1}. Indeed, the quotienting by the ideal
  generated by $Z$ gives the Lie algebra \eqref{eq:branch-1-1} with
  $\gamma \in (-1,1)$.
\item If $t_1 = 0$, then $t_3 = t_4 = 0$, so that the deformation is
  \begin{equation}
    \varphi = t_2 (c_7 - c_{10} + c_{14}) + c_3 - c_6 = t_2 (\zeta
    \beta B - \zeta \pi P + \beta\pi R) + \eta \beta B - \eta\pi P,
  \end{equation}
  with brackets
  \begin{equation}
    [H,\B] = \B \qquad [H,\P] = -\P \qquad [Z,\B] = t_2 \B \qquad
    [Z,\P] = -t_2 \P \qquad [\B,\P] = Z + t_2 R.
  \end{equation}
  Notice that $Z - t_2 H$ is central.  We must distinguish between two
  cases, depending on whether or not $t_2 = 0$:
  \begin{enumerate}
  \item if $t_2 \neq 0$, then we let $Z \mapsto t_2^{-1} Z$ and $H
    \mapsto H - t_2^{-1}Z$ and rescaling either $\B$ or $\P$ we can
    essentially set $t_2=1$ and arrive at the Lie algebra given in
    \eqref{eq:0-spacelike}.

  \item if $t_2 = 0$, then we have
    \begin{equation}
      \label{eq:spacelike-2-2}
      \boxed{[H,\B] = \B \qquad [H,\P] = -\P \qquad\text{and}\qquad
        [\B,\P] = Z,}
    \end{equation}
    which is isomorphic to a central extension of the Lie algebra
    \eqref{eq:branch-1-1} with $\gamma = -1$; that is, to a central
    extension of the lorentzian Newton Lie algebra.
  \end{enumerate}

\item If $t_1 =1$, then $t_2 = t_3 = 0$ and the deformation is
  \begin{equation}
    \varphi = 2 c_2 + 2 c_3 + t_4 (c_9 + c_{15}) = 2 \eta \zeta Z + 2
    \eta\beta B + t_4 (\zeta\pi B + \tfrac12 \pi\pi R),
  \end{equation}
  with brackets
  \begin{equation}
    [H,Z] = 2Z \qquad [H,\B] = 2 \B \qquad [Z,\P] = t_4 \B \qquad
    [\P,\P] = t_4 \R.
  \end{equation}
  We must distinguish between two cases, depending on whether or not
  $t_4 = 0$:
  \begin{enumerate}
  \item if $t_4 = 0$, then, rescaling $H$, we arrive at
    \begin{equation}
      [H,Z] = Z \qquad [H,\B] = \B \qquad\text{and}\qquad [\B,\P] = Z,
    \end{equation}
    which is isomorphic to \eqref{eq:spacelike-1} for $\gamma = 0$.
  \item and if $t_4 \neq 0$, then introducing $\varepsilon :=
    t_4/|t_4|$, we can bring the brackets to the following normal
    form:
    \begin{equation}
      \label{eq:spacelike-3-2}
      \boxed{[H,Z] = Z \qquad [H,\B] = \B \qquad [Z,\P] =
        \varepsilon \B \qquad [\P, \P] = \varepsilon \R
        \qquad\text{and}\qquad [\B,\P] = Z.}
    \end{equation}
    These Lie algebras are isomorphic to the extension of the
    euclidean or Poincaré Lie algebras by the dilatation $H$; that is,
    they are isomorphic to $\co(D,1) \ltimes \RR^{D,1}$ or $\co(D+1)
    \ltimes \RR^{D+1}$.
  \end{enumerate}

\item If $t_1=-1$, then $t_2 = t_4 = 0$ and the deformation is
  \begin{equation}
    \varphi = -2 c_2 - 2 c_6 + t_3 (c_8 - c_{13}) = -2 \eta\zeta Z - 2
    \eta\pi P + t_3 (\zeta\beta P - \tfrac12 \beta\beta R),
  \end{equation}
  with brackets
  \begin{equation}
    [H,Z] = -2 Z \qquad [H,\P] = -2 \P \qquad [Z,\B] = t_3 \P \qquad
    [\B,\B] = - t_3 \R.
  \end{equation}
  Exchanging $\B$ and $\P$ and changing the signs of $H$ and $Z$, we
  see that this case leads to isomorphic algebras to the case $t_1 = 1$.
\end{enumerate}

\subsubsection{Lightlike orbit branches}
\label{sec:lightl-orbit-branch}

Here $t_5 = t_6 = 0$ and $t_7 = 1$.  The system \eqref{eq:quadrics-ce}
of quadrics now sets $t_2 = t_3 = 0$ and imposes $t_1 t_4 = 0$, which
gives rise to two branches, depending on whether or not $t_1 = 0$:
\begin{enumerate}
\item $t_1 = 0$: in this case the deformation is
  \begin{equation}
    \varphi = t_4 (c_9 + c_{15}) + c_5 = t_4 (\zeta\pi B + \tfrac12
    \pi\pi R) + \eta \pi B,
  \end{equation}
  so that the brackets are
  \begin{equation}
    [Z,\P] = t_4 \B \qquad [H,\P] = \B \qquad\text{and}\qquad [\P,\P]
    = t_4 \R.
  \end{equation}
  We notice that $Z - t_4 H$ is central, so this deformation is a
  (possibly trivial) central extension.  We must distinguish between
  two cases, according to whether or not $t_4 = 0$.
  \begin{enumerate}
  \item If $t_4 = 0$, then we obtain
    \begin{equation}
      \label{eq:bargmann}
      \boxed{[H,\P] = \B \qquad\text{and}\qquad [\B, \P] = Z,}
    \end{equation}
    which is isomorphic to the \textbf{Bargmann algebra}: the
    universal central extension of the galilean algebra
    \eqref{eq:galilean}.

  \item If $t_4 \neq 0$ and introducing $\varepsilon = t_4/|t_4|$, we
    may redefine generators to arrive at a Lie algebra isomorphic to
    \eqref{eq:0-lightlike}.
  \end{enumerate}

\item $t_1 \neq 0$: in this case $t_4 = 0$ and the deformation is
  \begin{equation}
    \varphi = t_1 (2 c_2 + c_3 + c_6) + c_5 = t_1 (2 \eta\zeta Z +
    \eta\beta B + \eta\pi P) + \eta\pi B,
  \end{equation}
  so that the brackets are
  \begin{equation}
    [H,Z] = 2 t_1 Z \qquad [H,\B] = t_1 \B \qquad\text{and}\qquad
    [H,\P] = \B + t_1 \P.
  \end{equation}
  We can actually absorb $t_1$ into a redefinition of the generators
  and arrive at
  \begin{equation}
    \label{eq:lightlike-2}
    \boxed{[H,\B] = \B \qquad [H,\P] = \B + \P \qquad [H, Z] = 2 Z
      \qquad\text{and}\qquad [\B,\P] = Z.}
  \end{equation}
  This Lie algebra is isomorphic to a non-central extension of the Lie
  algebra \eqref{eq:branch-1-2}, which we recover quotienting by the
  ideal generated by $Z$.

\end{enumerate}

\subsubsection{Timelike orbit branches}
\label{sec:timel-orbit-branch}

In this case $t_5 = 0$, $t_6 =1$ and $t_7 = -1$.  The system
\eqref{eq:quadrics-ce} of quadrics implies that $t_2 = 0$, $t_4= -
t_3$ and $t_1 t_3 = 0$.  This then gives rise to two branches,
depending on whether or not $t_1 = 0$:
\begin{enumerate}
\item $t_1 = 0$: in this case the deformation is
  \begin{equation}
    \varphi = t_3 (c_8 - c_9 - c_{13} - c_{15}) + c_4 - c_5 = t_3
    (\zeta\beta P - \zeta \pi B - \tfrac12 \beta\beta R - \tfrac12
    \pi\pi R) + \eta \beta P - \eta \pi B,
  \end{equation}
  with brackets
  \begin{equation}
    [Z,\B] = t_3 \P \qquad [Z,\P] = - t_3 \B \qquad [\B,\B] = -t_3 \R
    \qquad [\P,\P] = - t_3 \R \qquad [H,\B] = \P \qquad [H,\P] = -\B.
  \end{equation}
  We notice that $Z-t_3 H$ is central for all $t_3$, so these Lie
  algebras are (possibly trivial) central extensions.  We
  distinguish between two cases depending on whether or not $t_3 = 0$.
  \begin{enumerate}
  \item If $t_3 = 0$, then we obtain
    \begin{equation}
      \label{eq:timelike-1-1}
      \boxed{[H,\B] = \P \qquad [H,\P] = -\B \qquad\text{and}\qquad
        [\B,\P] = Z.}
    \end{equation}
    This Lie algebra can be interpreted as the central extension (with
    central element $Z$) of the Lie algebra \eqref{eq:branch-1-3} with
    $\alpha = 0$; that is, a central extension of the euclidean Newton
    Lie algebra.

  \item If $t_3 \neq 0$, then introducing $\varepsilon := -
    t_3/|t_3|$, we can rescale generators to arrive at a Lie algebra
    isomorphic to \eqref{eq:0-timelike}.
  \end{enumerate}

\item $t_1 \neq 0$, so that $t_3 = t_4 = 0$, and the deformation is
  \begin{equation}
    \varphi = t_1 (2 c_2 + c_3 + c_6) + c_4 - c_5 = t_1 (2 \eta\zeta Z
    + \eta\beta B + \eta\pi P) +  \eta \beta P - \eta \pi B
  \end{equation}
  with brackets
  \begin{equation}
    [H,Z] = 2 t_1 Z \qquad [H,\B] = t_1 \B + \P \qquad [H,\P] = t_1 \P
    - \B.
  \end{equation}
  Here without loss of generality we can take $t_1 = \alpha > 0$
  and arrive at
  \begin{equation}
    \label{eq:timelike-2}
    \boxed{[H,\B] = \alpha \B + \P \qquad [H,\P] = -\B + \alpha \P
      \qquad [H,Z] = 2 \alpha Z \qquad\text{and}\qquad [\B,\P] = Z.}
  \end{equation}
  This is a non-central extension of the Lie
  algebra~\eqref{eq:branch-1-3} (for $\alpha > 0$) by the element
  $Z$.
\end{enumerate}

\subsection{Invariant inner products}
\label{sec:invar-inner-prod-1}

We shall now investigate the existence of invariant inner products on
the Lie algebras determined in this section, as we did in
Section~\ref{sec:invar-inner-prod} for the kinematical Lie algebras
classified in Section~\ref{sec:deform-stat-kinem-1}.  We shall prove
that only the trivial central extensions of the simple kinematical Lie
algebras $\so(D+2)$, $\so(D+1,1)$ and $\so(D,2)$ admit invariant inner
products.  To prove that the other Lie algebras in
Table~\ref{tab:ce-summary} do not admit such inner products, we shall
exploit the associativity condition~\eqref{eq:assoc}.  One of the
immediate consequences of this condition is that for a Lie algebra
$\g$ with an invariant inner product, $\g' = Z(\g)^\perp$, where
$Z(\g)$ is the centre and $\g' = [\g,\g]$ is the first derived ideal.
Therefore if $\g$ is such that $Z(\g) =0$ but $\g' \subsetneq \g$,
then $\g$ cannot admit an invariant inner product.  This is precisely
the situation of the Lie algebras in the bottom third (below the line) of
Table~\ref{tab:ce-summary}.

The first Lie algebra in Table~\ref{tab:ce-summary} (with brackets given by
\eqref{eq:cex-static}) does not admit an invariant inner
product.  Indeed, if $(-,-)$ is an associative symmetric bilinear
form, it follows that
\begin{equation}
  \delta_{ab} (Z,Z) = ([B_a,P_b],Z) = (B_a, [P_b,Z]) = 0
\end{equation}
and
\begin{equation}
  \delta_{ab} (Z,H) = ([B_a,P_b],H) = (B_a, [P_b,H]) = 0,
\end{equation}
so that $(Z,-) = 0$.  The exact same calculation shows that in the
Bargmann algebra \eqref{eq:bargmann} any associative symmetric
bilinear form has $(Z,-) = 0$.  A very similar argument shows that the
trivial central extensions of the euclidean
and Poincaré algebras \eqref{eq:0-lightlike} do not admit invariant
inner products either.  Indeed, if $(-,-)$ is any associative
symmetric bilinear form, then
\begin{equation}
  \delta_{ab} (H,H) = ([B_a,P_b],H) = (B_a, [P_b,H]) = 0
\end{equation}
and
\begin{equation}
  \delta_{ab} (H,Z) = ([B_a,P_b],Z) = (B_a, [P_b,Z]) = 0,
\end{equation}
so that $(H,-) = 0$.
The trivial central extensions of $\so(D+1,1)$, $\so(D+2)$ and $\so(D,2)$
do admit invariant inner products by taking the Killing form on the
simple factor and some non-zero value for $(Z,Z)$.

Finally, we treat the centrally extended Newton algebras.  The two
cases are very similar, so we give details only for the case of the
lorentzian algebra \eqref{eq:spacelike-2-2}.  Let $(-,-)$ be an
associative symmetric bilinear form.  We show that $(B_a,-) = 0$,
so that it is degenerate.  First of all, by rotational invariance,
$(B_a, H) = (B_a, Z) = 0$.  Let us calculate the others (in
abbreviated notation)
\begin{equation}
  \begin{split}
    (\B, \B) &= ([H,\B], \B) = (H, [\B,\B]) = 0\\
    (\B,\P) &= ([\R,\B],\P) = (\R, [\B, \P]) = (\R, Z) = 0.
  \end{split}
\end{equation}
The euclidean case \eqref{eq:timelike-1-1} is similar.  In summary,
only the trivial central extensions of the simple kinematical Lie
algebras $\so(D+1,1)$, $\so(D+2)$ and $\so(D,2)$ admit invariant inner
products.

\section{Deformations of the centrally extended static kinematical Lie algebra with $D=4$}
\label{sec:deform-centr-extend-4}

In this section $\tg$ denotes the universal central
extension of the static kinematical Lie algebra for $D=4$.  As in the
non-centrally extended case, $D=4$ is slightly more complicated due to
the semisimplicity of the rotation subalgebra $\so(4)\cong \so(3)
\oplus \so(3)$.  The notation is as in
Section~\ref{sec:kinem-lie-algebr}, in particular we shall let
$R_{ab}^\pm := \tfrac12 \left( R_{ab} \pm \tfrac12 \epsilon_{abcd}
  R_{cd}\right)$ span $\s \cong \so(4)$ and $B_a,P_a,H,Z$ span the
ideal $\th$ of $\tg$.  As usual we choose the canonical
dual basis $\beta_a, \pi_a, \eta,\zeta$ for $\th^*$.

The non-zero Lie brackets in that basis are given by
\begin{equation}
  \begin{split}
    [R^\pm_{ab}, R^\pm_{cd}] &= [R_{ab}, R_{cd}]^\pm\\
    [R^\pm_{ab}, B_c] &= \tfrac12 \left(\delta_{bc} B_a - \delta_{ac}
      B_b \mp \epsilon_{abcd} B_d \right)\\
    [R^\pm_{ab}, P_c] &= \tfrac12 \left(\delta_{bc} P_a - \delta_{ac}
      P_b \mp \epsilon_{abcd} P_d \right)\\
    [B_a, P_b] &= \delta_{ab} Z.
  \end{split}
\end{equation}

\subsection{The Chevalley--Eilenberg complex}
\label{sec:chev-eilenb-compl-4}

We apply the Hochschild--Serre decomposition theorem to calculate the
infinitesimal deformations from the $\s$-invariant subcomplex
$C^\bullet := C^\bullet(\th;\tg)^\s$.

We now proceed to enumerate bases for the spaces of cochains, noting
that $C^0$ is spanned by $H$ and $Z$.  The dimensions of $C^1$, $C^2$ and
$C^3$ are 8, 18 and 32, respectively, as can be checked using a roots
and weights calculation.  Natural bases are tabulated below.

\begin{table}[h!]
  \centering
  \caption{Basis for $C^1(\th;\tg)^{\s}$}
  \label{tab:basis-CE-4-C1}
  \begin{tabular}{*{8}{>{$}c<{$}}}
    a_1& a_2 & a_3 & a_4 & a_5 & a_6 & a_7 & a_8 \\\hline
    \eta H & \eta Z & \zeta H & \zeta Z & \beta B & \beta P & \pi B & \pi P\\
  \end{tabular}
\end{table}

\begin{table}[h!]
  \centering
  \caption{Basis for $C^2(\th;\tg)^{\s}$}
  \label{tab:basis-CE-4-C2}
  \begin{tabular}{*{9}{>{$}c<{$}}}
    c_1 & c_2 & c_3 & c_4 & c_5 & c_6 & c_7 & c_8 & c_9 \\\hline
    \eta\zeta H & \eta\zeta Z & \eta\beta B & \eta\beta P & \eta\pi B & \eta\pi P & \zeta\beta B & \zeta \beta P & \zeta \pi B\\[10pt]
    c_{10} & c_{11} & c_{12} & c_{13} & c_{14} & c_{15} & c_{16} & c_{17} & c_{18} \\\hline
    \zeta\pi P & \beta\pi H & \beta \pi Z & \tfrac12\beta\beta R^+ &\tfrac12\beta\beta R^- & \beta\pi R^+ & \beta\pi R^- & \tfrac12\pi\pi R^+& \tfrac12\pi\pi R^-\\
  \end{tabular}
\end{table}

\begin{table}[h!]
  \centering
  \caption{Basis for $C^3(\th;\tg)^{\s}$}
  \label{tab:basis-CE-4-C3}
  \begin{tabular}{*{8}{>{$}c<{$}}}
    b_1 & b_2 & b_3 & b_4 & b_5 & b_6 & b_7 & b_8 \\\hline
    \eta\zeta\beta B & \eta\zeta\beta P & \eta\zeta\pi B & \eta\zeta\pi P & \eta\beta\pi H & \eta\beta\pi Z & \eta\beta\beta R^+ & \eta\beta\beta R^-\\[10pt]
    b_9 & b_{10} & b_{11} & b_{12} & b_{13} & b_{14} & b_{15} & b_{16}\\\hline
    \eta\beta\pi R^+ & \eta\beta\pi R^- & \eta\pi\pi R^+ & \eta\pi\pi R^- & \zeta\beta\pi H & \zeta\beta\pi Z & \zeta\beta\beta R^+ & \zeta\beta\beta R^-\\[10pt]
    b_{17} &  b_{18} &  b_{19} &  b_{20} & b_{21} & b_{22} & b_{23} & b_{24} \\\hline
    \zeta\beta\pi R^+ & \zeta\beta\pi R^- & \zeta\pi\pi R^+ & \zeta\pi\pi R^- & \beta\pi\beta B & \beta\pi\beta P & \beta\pi\pi B & \beta\pi\pi P \\[10pt]
    b_{25} &  b_{26} &  b_{27} &  b_{28} & b_{29} & b_{30} & b_{31} & b_{32} \\\hline
    \epsilon\beta\beta\beta R^+ &  \epsilon\beta\beta\beta R^- & \epsilon\beta\beta\pi R^+ & \epsilon\beta\beta\pi R^- & \epsilon\beta\pi\pi R^+ & \epsilon\beta\pi\pi R^- & \epsilon\pi\pi\pi R^+  & \epsilon\pi\pi\pi R^-   \\
  \end{tabular}
\end{table}

The Chevalley--Eilenberg differential $\d : C^p \to C^{p+1}$ is
defined on generators by
\begin{equation}
  \d \eta = \d \beta_a = \d \pi_a = \d H = \d Z = 0
\end{equation}
and
\begin{equation}
  \begin{split}
  \d R_{ab}^\pm &= \tfrac12 \left(\beta_a B_b - \beta_b B_a + \pi_a P_b
    - \pi_b P_a \pm \epsilon_{abcd} (\beta_c B_d + \pi_c P_d) \right)\\
  \d B_a &= - \pi_a Z\\
  \d P_a &= \beta_a Z\\
  \d \zeta &= -\beta_a \pi_a.
\end{split}
\end{equation}

The Chevalley--Eilenberg differential $\d: C^1 \to C^2$ is given on
the basis by
\begin{equation}
  \d a_1 = \d a_2 = \d a_6 = \d a_7 = 0, \qquad \d a_5 = \d a_8 =
  c_{12}, \qquad \d a_4 = -c_{12} \qquad\text{and}\qquad \d a_3 = -c_{11},
\end{equation}
from where we see that $B^2= \RR\left<c_{11}, c_{12}\right>$.  The
differential $\d : C^2 \to C^3$ is given on the basis by
\begin{equation}
  \label{eq:CE-d-4}
  \begin{aligned}[m]
    \d c_1 & = b_5\\
    \d c_2 & = b_6\\
    \d c_3 & = -b_6\\
    \d c_4 & = 0\\
    \d c_5 & = 0\\
    \d c_6 & = -b_6\\
  \end{aligned}
  \qquad\qquad
  \begin{aligned}[m]
    \d c_7 & = -b_{14} - b_{21}\\
    \d c_8 & = -b_{22}\\
    \d c_9 & = -b_{23}\\
    \d c_{10} & = -b_{14} - b_{24}\\
    \d c_{11} & = 0\\
    \d c_{12} & = 0\\
  \end{aligned}
  \qquad\qquad
  \begin{aligned}[m]
    \d c_{13} & = -\tfrac12 b_{22} + \tfrac14 (b_{25} + b_{28})\\
    \d c_{14} & = -\tfrac12 b_{22} - \tfrac14 (b_{25} + b_{28})\\
    \d c_{15} & = \tfrac12 (b_{21} - b_{24} + b_{27} + b_{30})\\
    \d c_{16} & = \tfrac12 (b_{21} - b_{24} - b_{27} - b_{30})\\
    \d c_{17} & = -\tfrac12 b_{23} + \tfrac14 (b_{29} + b_{32})\\
    \d c_{18} & = -\tfrac12 b_{23} - \tfrac14 (b_{29} + b_{32})\\
  \end{aligned}
\end{equation}

The last piece of data that we need is the Nijenhuis--Richardson
bracket $[\![-,-]\!]: C^2 \times C^2 \to C^3$.
Table~\ref{tab:NR-dot-CE-4} collects the calculations of
$c_i \bullet c_j$, from where we can read off $[\![c_i, c_j]\!]$ by
symmetrisation.

\begin{table}[h!]\small
  \setlength{\extrarowheight}{4pt}
  \setlength{\tabcolsep}{3pt}
  \centering
  \caption{Nijenhuis--Richardson $\bullet$}
  \label{tab:NR-dot-CE-4}
  \begin{tabular}{>{$}c<{$}|*{18}{>{$}c<{$}}}
    \bullet & c_1 & c_2 & c_3 & c_4 & c_5 & c_6 & c_7 & c_8 & c_9 & c_{10} & c_{11} & c_{12} & c_{13} & c_{14} & c_{15} & c_{16} & c_{17} & c_{18} \\\hline
    c_1 & \zero & \zero & b_{1} & b_{2} & b_{3} & b_{4} & \zero & \zero & \zero & \zero & \zero & \zero & \zero & \zero & \zero & \zero & \zero & \zero \\
     c_2 & \zero & \zero & \zero & \zero & \zero & \zero & b_{1} & b_{2} & b_{3} & b_{4} & \zero & \zero & \zero & \zero & \zero & \zero & \zero & \zero \\
     c_3 & \zero & \zero & \zero & \zero & \zero & \zero & b_{1} & b_{2} & \zero & \zero & b_{5} & b_{6} & b_{7} & b_{8} & b_{9} & b_{10} & \zero & \zero \\
     c_4 & \zero & \zero & \zero & \zero & \zero & \zero & \zero & \zero & b_{1} & b_{2} & \zero & \zero & \zero & \zero & b_{7} & b_{8} & b_{9} & b_{10} \\
     c_5 & \zero & \zero & \zero & \zero & \zero & \zero & b_{3} & b_{4} & \zero & \zero & \zero & \zero & b_{9} & b_{10} & b_{11} & b_{12} & \zero & \zero \\
     c_6 & \zero & \zero & \zero & \zero & \zero & \zero & \zero & \zero & b_{3} & b_{4} & b_{5} & b_{6} & \zero & \zero & b_{9} & b_{10} & b_{11} & b_{12} \\
     c_7 & \zero & \zero & -b_{1} & -b_{2} & \zero & \zero & \zero & \zero & \zero & \zero & b_{5} & b_{6} & b_{15} & b_{16} & b_{17} & b_{18} & \zero & \zero \\
     c_8 & \zero & \zero & \zero & \zero & -b_{1} & -b_{2} & \zero & \zero & \zero & \zero & \zero & \zero & \zero & \zero & b_{15} & b_{16} & b_{17} & b_{18} \\
     c_9 & \zero & \zero & -b_{3} & -b_{4} & \zero & \zero & \zero & \zero & \zero & \zero & \zero & \zero & b_{17} & b_{18} & b_{19} & b_{20} & \zero & \zero \\
     c_{10} & \zero & \zero & \zero & \zero & -b_{3} & -b_{4} & \zero & \zero & \zero & \zero & b_{13} & b_{14} & \zero & \zero & b_{17} & b_{18} & b_{19} & b_{20} \\ 
    c_{11} & b_{13} & b_{14} & b_{21} & b_{22} & b_{23} & b_{24} & \zero & \zero & \zero & \zero & \zero & \zero & \zero & \zero & \zero & \zero & \zero & \zero \\
     c_{12} & -b_{5} & -b_{6} & \zero & \zero & \zero & \zero & b_{21} & b_{22} & b_{23} & b_{24} & \zero & \zero & \zero & \zero & \zero & \zero & \zero & \zero \\
     c_{13} &\zero & \zero & \zero & \zero & \zero & \zero & \zero & \zero & \zero & \zero & \zero & \zero & \zero & \zero & \zero & \zero & \zero & \zero \\
     c_{14} & \zero & \zero & \zero & \zero & \zero & \zero & \zero & \zero & \zero & \zero & \zero & \zero & \zero & \zero & \zero & \zero & \zero & \zero \\
     c_{15} &\zero & \zero & \zero & \zero & \zero & \zero & \zero & \zero & \zero & \zero & \zero & \zero & \zero & \zero & \zero & \zero & \zero & \zero \\
     c_{16} & \zero & \zero & \zero & \zero & \zero & \zero & \zero & \zero & \zero & \zero & \zero & \zero & \zero & \zero & \zero & \zero & \zero & \zero \\
     c_{17} &\zero & \zero & \zero & \zero & \zero & \zero & \zero & \zero & \zero & \zero & \zero & \zero & \zero & \zero & \zero & \zero & \zero & \zero \\
     c_{18} & \zero & \zero & \zero & \zero & \zero & \zero & \zero & \zero & \zero & \zero & \zero & \zero & \zero & \zero & \zero & \zero & \zero & \zero \\
   \end{tabular}
\end{table}

\subsection{Infinitesimal deformations}
\label{sec:infin-deform-CE-4}

From the action of the Chevalley--Eilenberg differential $\d$ on
$C^2$, described in Equation~\eqref{eq:CE-d-4}, we find that
\begin{equation}
  Z^2 =  B^2 \oplus \RR \left<2 c_2 + c_3 + c_6\right> \oplus \RR\left< c_3-c_6, c_4, c_5\right>
  \oplus \left<c_7 - c_{10} + c_{15} + c_{16}, c_8 - c_{13} - c_{14}, c_9 + c_{17} + c_{18}\right>.
\end{equation}
Therefore the most general infinitesimal deformation can be parametrised as
\begin{multline}
  \label{eq:inf-def-ce-4}
  \varphi_1 = t_1 (2 c_2 + c_3 + c_6) + t_2 (c_7 - c_{10} + c_{15} + c_{16})\\ + t_3(c_8 - c_{13} - c_{14})
  + t_4 (c_9 + c_{17} + c_{18}) + t_5 (c_3-c_6) + t_6 c_4 + t_7 c_5.
\end{multline}

\subsection{Obstructions}
\label{sec:obstructions-2}

The first obstruction is the class in $H^3$ of
$\tfrac12 [\![\varphi_1, \varphi_1]\!] = \varphi_1 \bullet \varphi_1$,
which we can calculate from the explicit expression for the
Nijenhuis--Richardson $\bullet$ product given in
Table~\ref{tab:NR-dot-CE-4}.  Doing so, we find that that
$\varphi_1 \bullet \varphi_1$ vanishes in cohomology if and only if it
vanishes on the nose, and this happens if and only if the following
quadric equations hold:
\begin{equation}
  \label{eq:quads-ce-4}
  \begin{split}
    0&= 2 t_1 t_2 - t_3 t_7 + t_4 t_6 \\
    0&= t_1 t_3 + t_3 t_5 - t_2 t_6\\
    0&= t_1 t_4 - t_4 t_5 + t_2 t_7.
  \end{split}
\end{equation}
If they are satisfied and since $[\![\varphi_1,\varphi_1]\!] = 0$, we
can take $\varphi_2 =0$ and the deformation integrates already to
first order.

We now observe that since $R^+ + R^- = R$,
and the sums of cochains appearing in $\varphi_1$ are
\begin{equation}
    c_{13} + c_{14} = \tfrac12 \beta\beta R \qquad
    c_{15} + c_{16} = \beta \pi R \qquad\text{and}\qquad
    c_{17} + c_{18} = \tfrac12 \pi\pi R,
\end{equation}
the expression~\eqref{eq:inf-def-ce-4} above for
$\varphi_1$ coincides \emph{mutatis mutandis} with the one in
Equation~\eqref{eq:inf-def-ce}.  Furthermore the conditions
\eqref{eq:quads-ce-4} and \eqref{eq:quadrics-ce} are identical, so that
the rest of the analysis proceeds as in the case $D\geq 5$.  As in
that case, here too the only Lie algebras admitting an invariant inner
product are the trivial central extensions of the simple kinematical
Lie algebras.

\section{Summary and conclusions}
\label{sec:conclusions}

Deformation theory provides a powerful and systematic approach to
classifying Lie algebras. When the Lie algebras in question have a
``sizeable'' semisimple subalgebra, the calculations are particularly
tractable due to the Hochschild--Serre spectral sequence, which
guarantees that we can work with a quasi-isomorphic subcomplex of the
deformation complex which is typically much smaller.

In this paper we have applied these techniques to classify kinematical
Lie algebras in dimension $D+1$ for $D>3$ (up to isomorphism) by
classifying deformations of the static kinematical Lie algebra $\g$.
The Lie algebra $\g$ admits a universal central extension $\tg$ and we
have classified its deformations as well.  This gives rise to a number
of extensions (trivial, central and non-central) of deformations of
$\g$.

Let us summarise the results obtained in this paper.  First of all we
summarise the kinematical Lie algebras.  It is convenient to tabulate
them to ease comparison with the classical results for $D=3$ and also
with the results for $D=2$ \cite{TAJMFKinematical2D}.
Table~\ref{tab:summary} lists the isomorphism classes of kinematical
Lie algebras.  In some cases we have changed basis to bring the Lie
algebra to a more familiar form.  Comparing with Table~1 in
\cite{MR857383} or Table~1 in \cite{JMFKinematical3D}, we see that,
unsurprisingly, there are some kinematical Lie algebras in $D=3$ which
do not exist in $D\geq 4$.  Those additional $D=3$ Lie algebras owe 
their existence to the $\so(3)$-invariant vector product $\RR^3 \times
\RR^3 \to \RR^3$, which is absent in $D>3$.  (There is a vector
product in $\RR^7$, but it is not invariant under $\so(7)$ but only
under a $\g_2$ subalgebra.)  Comparing with Table~1 in
\cite{TAJMFKinematical2D}, we see that also in $D=2$ there are
additional kinematical Lie algebras which owe their existence this
time to the symplectic structure on $\RR^2$.  (There is a symplectic
structure on $\RR^D$ for any even $D$, but only for $D=2$ it is
$\so(D)$-invariant.)

\begin{table}[h!]\small
  \centering
  \caption{Kinematical Lie algebras in $D+1$ dimensions for $D>3$}
  \label{tab:summary}
  \begin{tabular}{l|*{5}{>{$}l<{$}}|l|c}
    \multicolumn{1}{c|}{Eq.} & \multicolumn{5}{c|}{Nonzero Lie brackets} & \multicolumn{1}{c|}{Comments} &\multicolumn{1}{c}{Metric?}\\\hline
    \ref{eq:static} & & & & & & static & \\
    \ref{eq:galilean} & [H,\B] = \P & & & & & galilean & \\
    \ref{eq:branch-1-1} & [H,\B] = \gamma \B & [H,\P] = \P & & & & $\gamma \in (-1,1]$ & \\
    \ref{eq:branch-1-1} & [H,\B] = - \B & [H,\P] = \P & & & & lorentzian Newton & \\
    \ref{eq:branch-1-3} & [H,\B] = \B + \P & [H, \P] = \P & & & &  & \\
    \ref{eq:branch-1-2} & [H,\B] = \alpha \B + \P & [H,\P] = \alpha \P - \B & & & & $\alpha > 0$ & \\
    \ref{eq:branch-1-2} & [H,\B] = \P & [H,\P] = - \B & & & & euclidean Newton & \\
    \ref{eq:carroll} & & & [\B,\P] = H & & & Carroll & \\
    \ref{eq:poincare} & & [H,\P] = \B & [\B,\P] = H & [\P,\P] = \R & & $\p$  & \\
    \ref{eq:poincare} & & [H,\P] = -\B & [\B,\P] = H & [\P,\P] = -\R & & $\e$  & \\
    \ref{eq:hyperbolic} & [H,\B] = \B & [H,\P] = -\P &  [\B,\P] = H+\R  & & & $\so(D+1,1)$ & \checkmark \\
    \ref{eq:sphere} & [H,\B] = -\P & [H,\P] = \B & [\B,\P] = H &  [\B,\B]= \R &  [\P,\P] = \R & $\so(D,2)$ & \checkmark \\
    \ref{eq:sphere} & [H,\B] = \P & [H,\P] = -\B & [\B,\P] = H &  [\B,\B]= - \R &  [\P,\P] = - \R & $\so(D+2)$ & \checkmark \\
  \end{tabular}
\end{table}

As shown in Section~\ref{sec:invar-inner-prod}, only the simple Lie
algebras in this list (i.e., $\so(D+1,1)$, $\so(D+2)$ and $\so(D,2)$) admit an
associative (i.e., $\ad$-invariant) inner product.  This is in sharp
contrast with $D\leq 3$, where there are a number of non-simple metric
kinematical Lie algebras.

Next we summarise the deformations of the central extension of the
static kinematical Lie algebra with $D\geq 4$.
Table~\ref{tab:ce-summary} lists the isomorphism classes of these 
deformations with an identifying comment as to their structure or
their name, when known.  All of these Lie algebras share the following
Lie brackets (in abbreviated notation):
\begin{equation}
  [\R,\R] = \R \qquad [\R,\B] = \B \qquad [\R, \P] = \P \qquad [\R,
  H] = 0  \qquad\text{and}\qquad [\R, Z] =0.
\end{equation}
In the table we only list any additional non-zero brackets.  In
some cases we have changed notations ($H$ for $Z$ and $\B$ for $\P$)
for the sake of uniformity.  The table is divided into three: the top
third consists of (non-trivial) central extensions, the middle third of
trivial central extensions and the bottom third of non-central
extensions of kinematical Lie algebras.

Comparing with Table~2 in \cite{JMFKinematical3D}, we see that contrary
to the deformations of the static kinematical Lie algebra (without
central extension), the results in $D=3$ are \emph{mutatis mutandis}
the same as $D>3$.  The similar classification in $D=2$ does not
exist: the universal central extension of the $D=2$ static kinematical
Lie algebra has five central generators and not just one.

\begin{table}[h!]\tiny
  \setlength{\tabcolsep}{3pt}
  \centering
  \caption{Deformations of the centrally extended static kinematical
    Lie algebra in $D>3$}
  \label{tab:ce-summary}
  \setlength{\extrarowheight}{2pt}
  \begin{tabular}{l|*{6}{>{$}l<{$}}|l|c}
    \multicolumn{1}{c|}{Eq.} & \multicolumn{6}{c|}{Nonzero Lie brackets} & \multicolumn{1}{c|}{Comments} & \multicolumn{1}{c}{Metric?}\\\hline
    \ref{eq:cex-static} & [\B,\P] = Z & & & & & & centrally extended static  & \\
    \ref{eq:spacelike-2-2} & [\B,\P] = Z & [H, \B] = \B & [H,\P] = -\P & & & & central extension of lorentzian Newton  & \\
    \ref{eq:timelike-1-1} & [\B,\P] = Z & [H, \B] = \P & [H,\P] = -\B & & & & central extension of euclidean Newton  & \\
    \ref{eq:bargmann} & [\B,\P] = Z & [H, \B] = -\P & & & & & Bargmann  & \\\hline
    \ref{eq:0-lightlike} & [\B,\P] = H & [H, \B] = \P & & & [\B,\B] = \R & & $\e\oplus\RR$ & \\
    \ref{eq:0-lightlike} & [\B,\P] = H & [H, \B] = - \P & & & [\B,\B] = - \R & & $\p\oplus\RR$ & \\
  \ref{eq:0-spacelike} & [\B,\P] = H + \R & [H, \B] = \B & [H, \P] = -\P & & & & $\so(D+1,1) \oplus \RR$ &  \checkmark\\
  \ref{eq:0-timelike} & [\B,\P] = H & [H,\B] = \P & [H, \P] = - \B & & [\B,\B] = \R & [\P, \P] = \R & $\so(D+2)\oplus \RR$ & \checkmark\\
  \ref{eq:0-timelike} & [\B,\P] =  H & [H,\B] = -\P & [H, \P] =  \B & & [\B,\B] = - \R & [\P, \P] = - \R & $\so(D,2)\oplus \RR$ &  \checkmark\\\hline
   \ref{eq:branch-0-1} & [\B,\P] = Z & [H, \B] = \B & [H,\P] = \P & [H,Z] = 2 Z& & &  & \\
   \ref{eq:spacelike-1} & [\B,\P] = Z & [H, \B] = \gamma\B & [H,\P] = \P & [H,Z] = (\gamma+1) Z& & & $\gamma \in (-1,1)$  & \\
   \ref{eq:lightlike-2} & [\B,\P] = Z & [H, \B] = \B & [H, \P] = \B + \P & [H,Z] = 2 Z & & &  & \\
   \ref{eq:timelike-2} & [\B,\P] = Z & [H, \B] = \alpha \B + \P & [H,\P] = -\B + \alpha \P & [H,Z] = 2\alpha Z& & & $\alpha > 0$  & \\
   \ref{eq:spacelike-3-2} & [\B,\P] = Z & [Z,\B] = - \P & [H,\P] = \P & [H, Z] = Z & [\B,\B] = \R & & $\co(D,1) \ltimes \RR^{D,1}$ & \\
   \ref{eq:spacelike-3-2} & [\B,\P] = Z & [Z,\B] = \P & [H,\P] = \P & [H, Z] = Z & [\B,\B] = - \R & & $\co(D+1)\ltimes \RR^{D+1}$ & \\
  \end{tabular}
\end{table}

As in the case of kinematical Lie algebras, the only Lie algebras in
this table which admit an invariant inner product are the trivial
central extensions of the simple kinematical Lie algebras.  This
agrees with the results for $D=3$ as well.  We suspect that the case
of $D=2$ will provide us with some non-simple metric Lie algebras, but
we have not classified those deformations yet.

\section*{Acknowledgments}
\label{sec:acknowledgments}

This research is partially supported by the grant ST/L000458/1
``Particle Theory at the Higgs Centre'' from the UK Science and
Technology Facilities Council.

\providecommand{\href}[2]{#2}\begingroup\raggedright\endgroup

% \bibliographystyle{utphys}
% % \bibliography{AdS3,Sugra,Geometry,Algebra}
% \bibliography{KLAHD}

\end{document}